\documentclass[11pt,preprint]{aastex}

\slugcomment{accepted for publication in ApJ}

\begin{document}

\title{
Angular Momentum Transport by MHD Turbulence in Accretion Disks: Gas
Pressure Dependence of the Saturation Level of the Magnetorotational
Instability}

\author{
Takayoshi Sano\altaffilmark{1,2},
Shu-ichiro Inutsuka\altaffilmark{3},
Neal J. Turner\altaffilmark{4,5},
and
James M. Stone\altaffilmark{1,6}}

\altaffiltext{1}{
Department of Applied Mathematics and Theoretical Physics, Centre for
Mathematical Sciences, University of Cambridge, Wilberforce Road,
Cambridge CB3 0WA, UK}
\altaffiltext{2}{
Institute of Laser Engineering, Osaka University, Suita, Osaka
565-0871, Japan; sano@ile.osaka-u.ac.jp}
\altaffiltext{3}{
Department of Physics, Kyoto University, Kyoto 606-8502, Japan}
\altaffiltext{4}{
Department of Physics, University of California, Santa Barbara, CA
93106-9530}
\altaffiltext{5}{
Mail Stop 169-506, Jet Propulsion Laboratory, 4800 Oak Grove Dr,
Pasadena, CA 91109-8099}
\altaffiltext{6}{
Department of Astrophysical Sciences, Peyton Hall, Princeton
University, Princeton, NJ 08544-1001}

\begin{abstract}
The saturation level of the magnetorotational instability (MRI) is
investigated using three-dimensional MHD simulations.  The shearing
box approximation is adopted and the vertical component of gravity is
ignored, so that the evolution of the MRI is followed in a small local
part of the disk.
We focus on the dependence of the saturation level of the stress on
the gas pressure, which is a key assumption in the standard $\alpha$
disk model.
From our numerical experiments it is found that there is a weak
power-law relation between the saturation level of the Maxwell stress
and the gas pressure in the nonlinear regime; the higher the gas
pressure, the larger the stress. 
Although the power-law index depends slightly on the initial field
geometry, the relationship between stress and gas pressure is
independent of the initial field strength, and is unaffected by 
Ohmic dissipation if the magnetic Reynolds number is at least 10.  
The relationship is the same in adiabatic calculations, where pressure
increases over time, and nearly-isothermal calculations, where
pressure varies little with time.  
Over the entire region of parameter space explored, turbulence driven
by the MRI has many characteristic ratios such as that of the Maxwell
stress to the magnetic pressure.
We also find the amplitudes of the spatial fluctuations in density and
the time variability in the stress are characterized by the ratio of
magnetic pressure to gas pressure in the nonlinear regime.
Our numerical results are qualitatively consistent with an idea that
the saturation level of the MRI is determined by a balance between the
growth of the MRI and the dissipation of the field through
reconnection.
The quantitative interpretation of the pressure-stress relation,
however, may require advances in the theoretical understanding of
non-steady magnetic reconnection.
\end{abstract}


\clearpage
\section{INTRODUCTION}

Most existing models of accretion disks are based on the
$\alpha$-prescription of Shakura \& Sunyaev (1973).  
In this picture, the physical nature of the accretion torque is
unspecified, and the radial-azimuthal component of the stress tensor,
$w_{r \phi}$, is assumed to be a constant, $\alpha$, times the
pressure. 
Angular momentum is supposed to be transported outward by an
``anomalous viscosity''.   
The viscosity required by observed evolution timescale is orders of
magnitude greater than that resulting from ordinary molecular
viscosity.  
Dimensional analysis suggests that the anomalous viscosity may be
represented by $\alpha c_s H$, where $c_s$ and $H$ are the sound speed
and the disk scale height.  
The advantage of this approach is that many of the uncertainties
regarding the accretion stress are confined in the single parameter
$\alpha$. 

A promising physical mechanism for anomalous viscosity is turbulence. 
How turbulence might be driven has been an open question for many
years.  
Keplerian disks satisfy Rayleigh's hydrodynamical stability criterion,
as the specific angular momentum increases monotonically outward.  
No clear means for locally generating and sustaining hydrodynamic
turbulence has been identified.  
When a magnetic field is present, however, the condition for stability
is that the angular velocity $\Omega$ increases outward (Chandrasekhar
1961).  
This condition is usually violated in accretion disks.  
The presence of a weak magnetic field leads to magnetorotational
instability (MRI; Balbus \& Hawley 1991; 1998), which initiates and
sustains MHD turbulence. 
Numerical simulations using several different methods have shown that 
Maxwell and Reynolds stresses in the turbulence transport a
significant amount of angular momentum outward  (e.g. Hawley, Gammie,
\& Balbus 1995; 1996; Matsumoto \& Tajima 1995; Brandenburg et
al. 1995). 
The Maxwell stress is a few times larger than the Reynolds stress.

In magnetized accretion disks, it is likely that the rate of angular
momentum transport and the value of the $\alpha$-parameter are
determined by the saturation amplitude of the turbulence resulting
from MRI.  
Numerical simulations indicate that the time- and
volume-averaged Maxwell stress, $-B_r B_{\phi} / 4 \pi$, in the
saturated state is usually proportional to the time- and
volume-averaged magnetic pressure, $B^2 / 8 \pi$ (e.g., Hawley et
al. 1995; 1996).  
Therefore it is important to determine how the
magnetic pressure depends on the gas pressure, so that we can relate
MHD calculations to the many existing studies that use the $\alpha$
prescription.  
The magnetic pressure is controlled by generation and
dissipation processes, and the rate of dissipation may vary with the
gas pressure.  
The main aim of this work is to study how stress
depends on gas pressure in turbulence driven by the MRI.

In the Shakura-Sunyaev picture, $\alpha$ is the ratio of the accretion 
stress to the gas pressure.  
The saturation level of the $\alpha$ parameter ranges from $10^{-3}$
-- $0.1$ in ideal MHD simulations (e.g., Hawley et al. 1995; 1996). 
However, the important questions of how the MRI saturates and which
physical quantities determine the saturation level have not yet been
resolved.   
The ultimate goal of this work is to understand the mechanism of
nonlinear saturation of the MRI, so that we may predict the saturation
level from a small number of parameters.  
The first attempts to measure the parameter dependence of the stress
numerically were made by Hawley et al. (1995; 1996).  
They found that the saturation level depends on the field strength, in
disks penetrated by uniform vertical fields.   
On the other hand, the final state is independent of the initial field
strength if there is no net magnetic flux in the system.   
A single initial gas pressure was used in all the calculations, so
that the pressure dependence of the saturation level was not
addressed.   
Here we extend the work of Hawley et al. (1995; 1996).  
We reexamine the predictor function, explicitly considering the
pressure dependence. 
The calculations are run for hundreds of orbits for accurate
estimation of saturation levels. 

The gas pressure dependence of the saturated stress remains unclear. 
No numerical experiment has explicitly examined the dependence.
However, there is indirect evidence that higher gas pressures may
enhance the Maxwell stress.   
Stone et al. (1996) carried out three-dimensional simulations of the
MRI in a vertically stratified disk using both adiabatic and
isothermal equations of state.   
In the adiabatic case, temperatures increase with time due to
dissipation of magnetic energy, while in the isothermal version, the
temperature is constant.  
The pressures in the nonlinear regime differ substantially between the
two.   
The saturation level in the adiabatic run is higher than in the
isothermal run with the same initial condition, suggesting that higher
pressures may contribute to higher saturation levels of the stress.  
However the calculations include stratification, so that buoyancy as
well as gas pressure may affect the saturated state.   
The effects of gas pressure are studied separately in the present
paper by using the unstratified shearing box approximation (Hawley et
al. 1995). 

Another advantage of unstratified local simulations is that the
evolution can be followed for many orbits.   
Turbulence driven by the MRI is fluctuating and chaotic (Winters,
Balbus, \& Hawley 2003), and averages must be taken over long
intervals for reliable estimates of the saturation level.  
Most of our simulations are integrated for a few hundred orbits, and
time averages are taken over more than 50 orbits.  
These periods are long compared with previous work using similar
calculations. 

The plan of this paper is as follows.  
The basic equations and initial conditions are described in \S~2.  
We make the local shearing box approximation, in which total energy
increases over time because of the radial boundary conditions.  
Energy transfer and thermalization in the shearing box are also
discussed briefly in \S~2. 
Numerical results are shown in \S~3.  
The parameters defining the initial condition in the local
approximation are the gas pressure, the geometry and strength of the
magnetic field, the equation of state, the size of the calculation
box, and the numerical resolution.   
The effects of all parameters must be understood to construct a
saturation predictor function.  
However in this paper, we concentrate on the gas pressure, magnetic
field, and equation of state. 
The remaining parameters will be investigated in subsequent work.  
The evolution of the MRI in disks with zero and non-zero net magnetic
flux is examined in \S\S~3.1 and 3.2, respectively.  
The effects of magnetic dissipation are discussed in \S~3.3, while
\S~3.4 is devoted to general characteristics of MHD turbulence driven
by the MRI.   
Time variability of the turbulence and the interpretation of our
numerical results are discussed in \S~4.  
Finally, \S~5 is a brief summary.

\section{NUMERICAL METHOD}

\subsection{Basic Equations and Numerical Scheme}

The shearing box approximation is used in our numerical simulations,
because we focus on the local behavior of the MRI in the simplest
representation of accretion disks.
The vertical gravity is ignored so that all the physical quantities
are initially uniform except for the sheared velocity in the azimuthal
direction.
The equations to be solved are
\begin{equation}
\frac{\partial \rho}{\partial t} + 
\mbox{\boldmath $v$} \cdot \nabla \rho = 
- \rho \nabla \cdot \mbox{\boldmath $v$} ~,
\label{eqn:eoc}
\end{equation}
\begin{equation}
\frac{\partial \mbox{\boldmath $v$}}{\partial t} + 
\mbox{\boldmath $v$} \cdot \nabla \mbox{\boldmath $v$} = 
- \frac{\nabla P}{\rho} 
+ \frac{\mbox{\boldmath $J$} \times \mbox{\boldmath $B$}}{c \rho} 
- 2 \mbox{\boldmath $\Omega$} \times \mbox{\boldmath $v$} 
+ 2 q \Omega^2 x \hat{\mbox{\boldmath $x$}} ~,
\label{eqn:eom}
\end{equation}
\begin{equation}
\frac{\partial \epsilon}{\partial t} + 
\mbox{\boldmath $v$} \cdot \nabla \epsilon = 
- \frac{P \nabla \cdot \mbox{\boldmath $v$}}{\rho} +
\frac{4 \pi \eta \mbox{\boldmath $J$}^2}{c^2 \rho} ~,
\label{eqn:ene}
\end{equation}
\begin{equation}
\frac{\partial \mbox{\boldmath $B$}}{\partial t} = 
\nabla \times \left( \mbox{\boldmath $v$} \times \mbox{\boldmath $B$} 
- \frac{4 \pi \eta \mbox{\boldmath $J$}}{c} 
\right) ~,
\label{eqn:ineq}
\end{equation}
where 
\begin{equation}
 \mbox{\boldmath $J$} = \frac{c}{4 \pi} 
\left( \nabla \times \mbox{\boldmath $B$} \right)
\end{equation}
is the current density and $\epsilon$ is the specific internal energy.
The basic equations are written in a local Cartesian frame of
reference ($x$, $y$, $z$) corotating with the disk at angular
frequency $\Omega$, where $x$ is oriented in the radial direction, $y$
is in the azimuthal direction, and $z$ is in the vertical direction.
The terms $- 2 \mbox{\boldmath $\Omega$} \times \mbox{\boldmath $v$}$
and $2 q \Omega^2 x$ in the equation of motion (\ref{eqn:eom}) are the
Coriolis force and the tidal expansion of the effective potential with
a constant $q = 3 / 2$ for a Keplerian disk, respectively. 
The gas is assumed to be ideal, with pressure $P = ( \gamma - 1 )
\rho \epsilon$ where $\gamma$ is the ratio of the specific heats.
In this paper, we examine both the ideal MHD and resistive MHD cases.
The induction equation (\ref{eqn:ineq}) includes a term for the Ohmic
dissipation, where $\eta$ is the magnetic diffusivity.
The energy equation (\ref{eqn:ene}) has the Joule heating term.

These equations are solved with the second-order Godunov-type scheme
developed by T. Sano \& S. Inutsuka (2004, in preparation).  
Operator splitting is used.
The hydrodynamical part of the equations is solved by a Godunov
method, using the exact solution of the simplified MHD Riemann
problem.   
The Riemann problem is simplified by including only the tangential
component of the field.  
The characteristic velocity is then that of the magneto-sonic wave
alone, and the MHD Riemann problem can be solved in a way similar to
the hydrodynamical one (Colella \& Woodward 1984).  
The piecewise linear distributions of flow quantities are calculated
with a monotonicity constraint following van Leer's (1979) method.  
The remaining terms, the magnetic tension component of the equation of
motion and the induction equation, are solved by the MoC-CT method
(Stone \& Norman 1992), guaranteeing $\nabla \cdot \mbox{\boldmath 
 $B$} = 0$ to within round-off error throughout the calculation (Evans
\& Hawley 1988).  

The accuracy of the scheme has been demonstrated by various test
problems and by calculations of the numerical growth rate of the MRI
(Sano 1998; Sano \& Stone 2002a).
The inclusion of the Ohmic dissipation term into the CT method is
straightforward.
For ideal MHD, the electro-motive force, $\mbox{\boldmath $v$}
\times \mbox{\boldmath $B$}$, is defined at the edge of each cell.
For resistive MHD, we evaluate the current density $\mbox{\boldmath
 $J$}$ at the same position, and calculate a new electro-motive force
including the dissipation term, $- \eta \mbox{\boldmath $J$}$.

The energy equation can be solved in either the total energy or the
internal energy form.  
In most cases we use the conservative, total energy form because it
allows a more complete analysis of the energy budget.  
In saturated turbulence driven by the MRI, the main source of
heating is magnetic dissipation (Sano \& Inutsuka 2001).  
Under non-conservative numerical schemes, magnetic energy can be lost
from the system, leading to time increases in gas pressure that are
slower than obtained using a total energy scheme.  
We use the internal energy scheme in a few models for comparing with
the results of the total energy scheme.

\subsection{Initial Conditions and Normalization}

In the initial equilibrium, the tidal force in the local effective
potential balances the Coriolis force, and both are much greater than
magnetic forces.
The relative importance of the field is given by the ratio between the
gas and magnetic pressures; $\beta = P / P_{{\rm mag}} = (2 / \gamma)
c_{s}^2 / v_{{\rm A}}^2$, where $c_{s} = ( \gamma P / \rho)^{1/2}$ and
$v_{{\rm A}} = B / (4 \pi \rho)^{1/2}$ are the sound velocity and
Alfv{\'e}n speed. 
The initial plasma beta $\beta_0$ is larger than 100 for all the
models shown in this paper.
The initial distribution of the azimuthal velocity is given by $v_{y0}
(x) = - q \Omega x$.
The uniform density and gas pressure are assumed to be $\rho_0$ and
$P_0$, respectively. 

Because the evolution of the MRI is sensitive to whether the net flux
of the vertical field is zero or finite, we consider two kinds of
initial field geometries: a uniform vertical field $B_z = B_0$ and a
field with zero net flux, $B_z(x) = B_0 \sin (2 \pi x / L_x )$.  Here
$L_x$ is the size of the shearing box in the radial direction.

The system of equations is normalized using the initial density
($\rho_0 = 1$) and the angular velocity ($\Omega = 10^{-3}$),
following Hawley et al. (1995).  However, lengths are normalized
differently from previous studies so that the gas pressure dependence
of the saturation level may be examined.  The local shearing box has
three possible scales of length: the pressure scale height $H_0 =
(2/\gamma)^{1/2} c_{s0} / \Omega$, the unstable wavelength of the MRI
$\lambda_0 = 2 \pi v_{{\rm A}0} / \Omega$, and the size of the box.
Here $c_{s0}$ and $v_{{\rm A}0}$ are the initial sound velocity and
Alfv{\'e}n speed.

While lengths in previous work are normalized by $H_0$, we choose the
vertical height of the box $L_z = 1$ as the unit length.  Then both
the gas pressure $P_0$ and the field strength $B_0$ are independent
parameters, and determine the ratios $H_0 / L_z$ and $\lambda_0 /
L_z$, respectively.  Note that the ratio of the box size to the disk
thickness $H_0$ depends on the choice of initial gas pressure $P_0$ in
this normalization.  The radial and azimuthal sizes of the shearing
box are taken to be $L_x = L_z$ and $L_y = 4 L_z$.
The primary goal of this paper is to understand the dependence of the
saturation level on physical quantities.  For this purpose, we perform
a few extreme models in which the box size exceeds the pressure scale
height of the disk, or the gas pressure increases by three orders of
magnitude from its initial value.  Care must be taken when applying
the results to real accretion disks.

Spatially uncorrelated perturbations in the gas pressure and azimuthal
velocity are imposed at the beginning of each calculation.
The fluctuations have a zero mean value with a maximum amplitude of
$| \delta P | / P_0 = \beta_0^{-1}$ and $| \delta v | / c_{s0} = 0.1
\beta_0^{-1/2}$.
The amplitude of the initial fluctuations is less than 1 \%, because
$\beta_0 \ge 100$.
Most of the calculations use a standard grid resolution of $32 \times
128 \times 32$ with uniform zoning.  
In the azimuthal and vertical direction, periodic boundary conditions
are used.
For the radial boundary, a sheared periodic boundary condition (Hawley
et al. 1995) is adopted.

\subsection{Heating in the Shearing Box}

In the shearing box, angular momentum is transported by Maxwell and
Reynolds stresses, with sum
\begin{equation}
w_{xy} = - \frac{B_x B_y}{4 \pi} + \rho v_x \delta v_y \;,
\label{eq:torque}
\end{equation}
where $\delta v_y \equiv v_y + q \Omega x$ is the deviation from the
background shear motion.
The stress $w_{xy}$ is closely related to the total
energy within the box defined as 
\begin{equation}
\Gamma \equiv 
\int dV \left[ 
\rho \left( \frac{v^2}{2} + \epsilon + \phi \right) +
\frac{B^2}{8\pi} 
\right] 
\label{eq:Etotal}
\end{equation}
(Hawley et al. 1995), where $\phi = - q \Omega^2 x^2$ is the tidal
expansion of the effective potential. 
Using the evolution equations for the resistive MHD system
[eqs.~(\ref{eqn:eoc}) -- (\ref{eqn:ineq})], the
time-derivative of the above equation gives  
\mbox{\boldmath $$}
\begin{eqnarray}
\frac{d \Gamma }{dt} &=&
- \int d \mbox{\boldmath $A$} \cdot 
\left\{ \rho \mbox{\boldmath $v$} 
\left[ \frac{v^2}{2} + \epsilon + \frac{P}{\rho} + \phi \right]
\right. \nonumber \\
& & \left. + \frac{1}{4\pi}
\left[ \mbox{\boldmath $B$} \times 
( \mbox{\boldmath $v$} \times \mbox{\boldmath $B$} )
- \eta \mbox{\boldmath $B$} \times 
( \nabla  \times \mbox{\boldmath $B$} ) \right] \right\} \nonumber \\
&=& q \Omega L_x \int_{X} dA 
\left( \rho v_x \delta v_y - \frac{B_x B_y}{4\pi} \right) \nonumber \\
&=& q \Omega L_x \int_{X} dA w_{xy} \;, 
\label{eq:dEtdt}
\end{eqnarray}
where $dA$ is the surface element and the integral is taken over
either of the radial boundaries.
Thus the rate of energy input through the sheared periodic radial
boundary is proportional to the stress $w_{xy}$ at the boundary.  Note
that the final expression of equation (\ref{eq:dEtdt}) does not
explicitly depend on the resistivity.

If the stress $w_{xy}$ at the boundary is positive, or the angular
momentum flux through the box is outward, the total energy of the
system increases.  The source of the injected energy is the background
shear motion.  In realistic disks, positive stresses lead to inward
mass accretion, bringing a loss of gravitational energy.  The gain in
total energy in the shearing box represents this energy release.  The
injected energy goes to the magnetic field due to the growth of MRI.
Magnetic energy is then thermalized via magnetic reconnection.

In saturated turbulence, the time-averaged magnetic and kinetic
energies are nearly constant.
Furthermore, the density varies little, so that the change in
potential energy $\rho \phi$ is negligible compared with the other
terms in equation (\ref{eq:Etotal}).
All the energy gain of the system is therefore finally deposited by
the thermal energy $E_{\rm th} = P / (\gamma - 1)$, and equation
(\ref{eq:dEtdt}) can be written as
\begin{equation}
\langle \negthinspace \langle \dot{E_{\rm th}} \rangle \negthinspace
\rangle = q \Omega \langle \negthinspace \langle w_{xy} \rangle
\negthinspace \rangle \;,
\label{eqn:edot}
\end{equation}
where the double brackets $\langle \negthinspace \langle \rangle
\negthinspace \rangle$ denote time- and volume-averaged quantities
\footnote{We also use the single brackets $\langle f \rangle$ to
denote a volume average of quantity $f$.}.
This {\em fluctuation-dissipation relation} has been clearly
demonstrated in numerical simulations (Sano \& Inutsuka 2001).
A similar relation holds in cylindrical coordinates for the global
disk problem (Balbus \& Papaloizou 1999).

If cooling processes are inefficient and the stress $\langle
\negthinspace \langle w_{xy} \rangle \negthinspace \rangle$ is
constant, the gas pressure increases linearly with time. 
The time evolution is approximately given by
\begin{equation}
\langle \negthinspace \langle P(t) \rangle \negthinspace \rangle 
= P_0 + q ( \gamma - 1 ) \Omega \langle \negthinspace \langle w_{xy}
\rangle \negthinspace \rangle t \;.
\label{eqn:pt}
\end{equation}
Assuming $q = 3/2$ and $\gamma = 5/3$, the plasma beta would evolve as
\begin{eqnarray}
\frac{\langle \negthinspace \langle P(t) \rangle \negthinspace
  \rangle}{\langle \negthinspace \langle P_{\rm mag} \rangle
  \negthinspace \rangle} &=& \frac{P_0}{\langle \negthinspace \langle
  P_{\rm mag} \rangle \negthinspace \rangle} 
+ \Omega \frac{\langle \negthinspace \langle w_{xy} \rangle
  \negthinspace \rangle}{\langle \negthinspace \langle P_{\rm mag}
  \rangle \negthinspace \rangle} t \nonumber \\ 
&\approx& \beta_0 \frac{P_{{\rm mag},0}}{\langle \negthinspace \langle
  P_{\rm mag} \rangle
  \negthinspace \rangle} + 300 \left( \frac{\langle \negthinspace
  \langle w_{xy} \rangle \negthinspace \rangle/\langle \negthinspace
  \langle P_{\rm mag} \rangle \negthinspace \rangle}{0.5} \right)
  \left( \frac{t / t_{\rm rot}}{100} \right) \;.
\label{eqn:beta}
\end{eqnarray}
Here we use a relation $\langle \negthinspace \langle w_{xy} \rangle
\negthinspace \rangle / \langle \negthinspace \langle P_{\rm mag}
\rangle \negthinspace \rangle \approx 0.5$ obtained in previous
numerical simulations (e.g., Hawley et al. 1995).  Equation
(\ref{eqn:beta}) indicates that after a hundred orbits, the plasma
beta must be larger than a few hundred, even though the first term on
the right hand side of equation (\ref{eqn:beta}) can be much smaller.
The change in gas pressure over 100 orbits can be large, and should be
taken into account in analyzing the effects of gas pressure.

In realistic situations, however, the gas pressure can be reduced by
radiative cooling or by expansion in the vertical direction.
Instead of including cooling processes, we examine the effects of
cooling simply by changing the ratio of the specific heats
$\gamma$.
For nearly isothermal simulations, we use $\gamma = 1.001$, while
$\gamma = 5/3$ is used for adiabatic simulations.
	
\section{RESULTS}

The nonlinear evolution of the MRI is investigated using
three-dimensional MHD simulations.
Various initial conditions are used to reveal the dependence of the
saturation amplitude on physical quantities such as the gas pressure.
Parameters for all the models are listed in Tables 1 -- 3.
Model names are shown in column 1.
The first letter in the model name denotes the initial field
geometry. 
The labels of those with zero net vertical flux contain the
letter S.
Models started with a uniform $B_z$ have a label beginning with Z.
The following two numbers, $N_1$ and $N_2$, indicate the initial field
strength and the initial plasma beta, respectively.
The field strength is given by $v_{{\rm A}0} = B_0 / ( 4 \pi \rho_0
)^{1/2} = 2^{(N_1 - 7)} \times 10^{-4}$ using the tens digit $N_1$, so
that the initial field is stronger as $N_1$ is larger.
The next digit, $N_2$, stands for the size of the initial plasma beta
($\beta_0 = 10^{2 N_2}$).
Among those with the same $N_1$, the gas pressure is initially higher
if $N_2$ is larger.

The suffixes indicate $\gamma = 1.001$ ``isothermal'' (i), $\gamma =
5$ (g), cases using the internal energy scheme (e), those started with
a localized field in a small part of the disk (p), and runs with a
uniform diffusivity (r) or an anomalous diffusivity (a).

Column 2 is the initial plasma beta $\beta_0$.
The initial gas pressure $P_0$ and Alfv{\'e}n speed $v_{{\rm A}0}$ are
listed in column 3 and 4.
The scale height $H_0$ (column 5) and characteristic wavelength of
the MRI $\lambda_0$ (column 6 in Tables 1 and 2) are calculated from
$P_0$ and $v_{{\rm A}0}$.
In Table 3, column 6 is the size of the magnetic diffusivity $\eta$.
For all the models, we use a shearing box of size $1 \times 4 \times
1$, and a grid of $32 \times 128 \times 32$ zones.  Each grid cell is
a cube of size $\Delta = 1/32$.
The effects of the box size and numerical resolution may
be important also, and are discussed in a subsequent paper.

The total evolution time in units of the rotation time is listed in
column 9.
In addition to model parameters, the tables include the saturation
levels of a few quantities for each model.
Time- and volume-averaged Maxwell and Reynolds stresses ($\langle
\negthinspace \langle w_M \rangle \negthinspace \rangle$ and $\langle
\negthinspace \langle w_R \rangle \negthinspace \rangle$) are listed
in column 10 and 11, respectively.
The time average is taken over the last 50 orbits of each calculation
and given in term of the initial gas pressure $P_0$.
The change in the gas pressure ($\langle \negthinspace \langle P
\rangle \negthinspace \rangle / P_0$) is listed in column 12.
The $\alpha$ parameter of Shakura \& Sunyaev is given by $\alpha = (
\langle \negthinspace \langle w_M \rangle \negthinspace \rangle +
\langle \negthinspace \langle w_R \rangle \negthinspace \rangle) /
\langle \negthinspace \langle P \rangle \negthinspace \rangle$, which
is listed in column 13.
Note that normalization is done using the gas pressure in the
nonlinear regime $\langle \negthinspace \langle P \rangle
\negthinspace \rangle$, and not the initial value $P_0$.

\subsection{Zero Net Flux $B_z$ Models}

\subsubsection{Gas Pressure Dependence (Fiducial Models)}

First, we consider the cases without net magnetic flux in the disk.
The initial field is purely vertical and has a sinusoidal distribution
with radius; $B_z(x) = B_0 \sin ( 2 \pi x / L_x)$.
The direction of the field is upward in $x < 0$ and downward in $x >
0$, and its average over the entire domain is zero.
The typical time evolution of ($a$) the magnetic energy, ($b$) the
gas pressure, ($c$) the Maxwell stress, and ($d$) the Reynolds stress are
shown in Figure~\ref{fig:ts14}, where time is measured in orbits $t_{\rm
  rot} = 2 \pi / \Omega$.
The parameters of this fiducial model (S52) are $\beta_0 = 10^4$ and
$P_0 = 3.125 \times 10^{-6}$.
The pressure scale height and the MRI wavelength are initially $H_0 =
(2/\gamma)^{1/2} c_{s0}/\Omega = 2.5$ and $\lambda_0 = 2 \pi v_{{\rm
A}0}/\Omega = 0.16$.
The calculation box size is $1 \times 4 \times 1$, so that the
vertical size corresponds to 2/5 of the scale height of the disk
$H_0$.

The magnetic energy is amplified by the exponential growth of the
instability during the first few orbits (see Fig.~\ref{fig:ts14}$a$).
Then MHD turbulence is initiated and sustained until the end of the
calculation at 600 orbits.
The initial magnetic field is purely vertical, and the magnetic energy
$\langle B_z^2 / 8 \pi \rangle / P_0 \approx 5 \times 10^{-5}$.
The saturated amplitude $\langle B^2
/ 8 \pi\rangle / P_0 \approx 5 \times 10^{-2}$ is greater by about
three orders of magnitude.
In the saturated turbulence, the azimuthal component of the magnetic
pressure dominates the other components by an order of magnitude.
The ratios of each component ($\langle \negthinspace \langle B_y^2
\rangle \negthinspace \rangle / \langle \negthinspace \langle B_z^2
\rangle \negthinspace \rangle \approx 20$ and $\langle \negthinspace
\langle B_x^2 \rangle \negthinspace \rangle / \langle \negthinspace
\langle B_z^2 \rangle \negthinspace \rangle \approx 3$) are nearly
constant during the turbulent phase.

The gas pressure, on the other hand, continues to increase throughout
the evolution (see Fig.~\ref{fig:ts14}$b$), because no cooling
processes are included in the energy equation.
The main source of heating is the dissipation of magnetic fields
(Sano \& Inutsuka 2001).
The volume-averaged gas pressure at 300 orbits, $\langle P \rangle /
P_0 \approx 12$, is much larger than the magnetic pressure in the
saturated turbulent state, $\langle P_{\rm mag} \rangle / P_0 \sim
0.01$.
The magnetic energy is almost saturated while the gas pressure is
increasing, and thus the ratio between the magnetic and gas pressure
$\beta = P / P_{\rm mag}$ increases with time throughout the
calculation.

The efficiency of angular momentum transport is given by the
radial-azimuthal ($x$-$y$) component of the stress tensor $w_{xy}$.
The time evolutions of the Maxwell ($w_{M} \equiv - B_{x} B_{y} / 4
\pi$) and Reynolds stress ($w_{R} \equiv \rho v_{x} \delta v_{y} $)
are shown in Figures~\ref{fig:ts14}$c$ and \ref{fig:ts14}$d$,
respectively.
As is shown in previous work (e.g., Hawley et al. 1995; 1996), the
Maxwell stress always dominates the Reynolds stress by a factor of
about 5.
The stress fluctuates with time, but the amplitude of
the time variation is much smaller than in cases started with a
uniform $B_z$ (Sano \& Inutsuka 2001; see \S~4.1).
Although the Maxwell and Reynolds stresses are nearly saturated in the
nonlinear regime, the long term evolution shows a gradual increase.
To clarify this trend, we take the time average of the Maxwell stress
every 50 orbits, depicted in Figure~\ref{fig:ts14}$c$ by circles.
A slight positive slope can be seen in the evolution of the
time-averaged Maxwell stress.
Hereafter, we focus on this gradual increase of the stress.

During the nonlinear turbulent phase, both the gas pressure and the
Maxwell stress increase with time.
Thus it may be interesting to examine the correlation between these
two quantities.
In Figure~\ref{fig:wmps1}$a$, the volume-averaged Maxwell stress is
plotted as a function of the volume-averaged gas pressure.
The models shown in this figure are S51, S52, and S53.
Parameters for the three models are identical except for the initial
gas pressure $P_0$.
 
Each model evolves toward the upper right on this diagram, because
both the gas pressure and stress increase with time. 
Note that the horizontal axis can be regarded as time.
The increase of the gas pressure is significant when the initial value
is low. 
For the lowest $P_0$ model (S51), the gas pressure at the end of the
calculation is about 3 orders of magnitude larger than $P_0$.
The evolutionary track of model S51 almost overlaps with that
of S52 in the later stages.
The gas pressure in the highest $P_0$ model (S53) is nearly constant
because the initial plasma beta is very large $\beta_0 = 10^6$ for
this model.
The saturation level of the Maxwell stress in S53 is time-independent,
and slightly higher than in the other two models.
Figure~\ref{fig:wmps1}$a$ shows that higher gas pressure is associated
with larger stress.

Figure~\ref{fig:wmps1}$b$ is the same diagram as
Figure~\ref{fig:wmps1}$a$, but the volume- and time-averaged
values are plotted instead of the volume-averages. 
Models S51, S52, and S53 are shown by circles, triangles, and squares,
respectively.
The time average is taken over every 50 orbits after 50 orbits.
Obviously a power-law relation between the gas pressure and stress 
appears.
The power-law index $q$ of $\langle \negthinspace \langle w_M \rangle
\negthinspace \rangle \propto \langle \negthinspace \langle P \rangle
\negthinspace \rangle^q$ is about one quarter.  That is, the Maxwell
stress is roughly proportional to $P^{1/4}$.

In previous work on the nonlinear development of the MRI, the initial
gas pressure is usually used to normalize the stress.  However, as
seen from Figure~\ref{fig:wmps1}, the gas pressure can increase by
many orders of magnitude during the nonlinear evolution, and the
stress depends on the gas pressure in the nonlinear regime.  Thus when
analyzing long-term calculations, it may be better to normalize the
stress by the gas pressure averaged over the same time interval.

The short-term variability of MRI-driven turbulence is chaotic
(Winters et al. 2003).  However, averaging over a longer period of 50
orbits reveals a power-law relation between $\langle \negthinspace
\langle P \rangle \negthinspace \rangle$ and $\langle \negthinspace
\langle w_{M} \rangle \negthinspace \rangle$.  To find how long an
averaging period is needed, we performed two versions of calculation
S52, using different spatial distributions of the initial
perturbations.  The time evolutions of the Maxwell stress differ.
However, the stress averaged over 50 orbit periods follows the same
track, shown by the dotted line in Figure~\ref{fig:wmps1}$b$.
Therefore, fifty orbits may be a long enough averaging period when
examining the long-term evolution.
 
\subsubsection{Independence of the Initial Field Strength}

The power-law relation between the gas pressure and stress is found to
be independent of the initial field strength.
Figure~\ref{fig:wmpbz} shows the saturation level of the Maxwell stress
including models started with initial fields of different strengths.
Models S41 -- S43 ($circles$) have $B_0$ half as strong as
the fiducial ones (S51 -- S53), which are shown by triangles. 
The squares (S61 -- S63) and crosses (S71 and S72) are
those with fields 2 and 4 times stronger than the fiducial runs.
The initial beta value is $10^2$, $10^4$, or $10^6$ for each $B_0$
case, so that the initial gas pressures differ
(see Table 1).
All the parameters other than $B_0$ and $P_0$ are identical for all
the models shown in Figure~\ref{fig:wmpbz}.
The time averages of the Maxwell stress are taken over the last 50
orbits of each calculation.

As seen from the figure, all the models follow the same evolutionary
track on this diagram.
The saturated stress is independent of both the initial field
strength $B_0$ and the initial gas pressure $P_0$.
This means that information about the initial conditions is lost
in the nonlinear regime.
The time-averaged stress can be fitted by a simple function of the
time-averaged gas pressure in the nonlinear regime $\langle
\negthinspace \langle P \rangle \negthinspace \rangle$.
The power-law index is about $1/4$ and the best fit is $q = 0.28$
among all the models in Figure~\ref{fig:wmpbz}.

Note that if the stress is plotted against the initial gas pressure,
then no clear correlation is seen.
The saturation level of the stress normalized by the initial
pressure widely ranges from $10^{-5}$ to unity (see Table 1).
The gas pressure at the end of the calculation is 2 -- 3 orders of
magnitude larger than $P_0$ for the cases with $\beta_0$ relatively
small.
Thus the choice of the normalization, $P_0$ or $\langle \negthinspace
\langle P \rangle \negthinspace \rangle$, makes a huge difference in
the normalized stress.
The last column in Table 1 lists the total stress divided by the
pressure in the nonlinear regime; $\alpha \equiv (\langle \negthinspace
\langle w_{M} \rangle \negthinspace \rangle + \langle \negthinspace
\langle w_{R} \rangle \negthinspace \rangle)/\langle \negthinspace
\langle P \rangle \negthinspace \rangle$.
The amplitude of the $\alpha$ parameter ranges typically from
$10^{-3}$ to $10^{-4}$ with this normalization.

\subsubsection{Effects of the Equation of State}

In this subsection, the effects of the equation of state are examined.
We assume $\gamma = 5/3$ in all the models shown in
Figure~\ref{fig:wmpbz}, and  thus the thermal energy and gas pressure
increase monotonically with time.
In realistic systems, however, cooling processes can modify the
temperature of the disks.
Instead of implementing cooling terms in numerical simulations, we
demonstrate the effects of cooling processes simply by changing the
ratio of the specific heats $\gamma$.
Nearly isothermal models with $\gamma = 1.001$ are listed in Table 1.
In general, the Maxwell stress in the ``isothermal'' run is smaller
than that in the adiabatic run.
For example, the saturation level of the Maxwell stress for model S52i
is $\langle \negthinspace \langle w_M \rangle \negthinspace \rangle /
P_0 = 0.0031$, which is less than half of that for the adiabatic
counterpart S52.
For model S52, the gas pressure at the end of the calculation is 30
times larger than $P_0$.
The difference in the stresses may be due to an effect of the enhanced
gas pressure.

The saturation levels of the Maxwell stress in models S52i, S53i, S62i,
and S63i are shown in Figure~\ref{fig:wmgpbz} as a function of the gas
pressure.
The gas pressure in these models is unchanged throughout the
calculation so that always $\langle \negthinspace \langle P \rangle
\negthinspace \rangle \approx P_0$. 
In contrast to the adiabatic models, the time-averaged stress in the
``isothermal'' models is found to be almost constant with time.
The time average is taken from 50 to 300 orbits, and the error bars in
the figure show the dispersions of time averages taken
every 50 orbits.
The dotted line in the figure is the $\langle \negthinspace \langle P
\rangle \negthinspace \rangle$-$\langle \negthinspace \langle w_{M}
\rangle \negthinspace \rangle$ relation obtained from the adiabatic
runs (Fig.~\ref{fig:wmpbz}).
The ``isothermal'' runs with the higher initial pressure $P_0$ have
larger saturation amplitude in the stress.  
Furthermore the relation between $\langle \negthinspace \langle w_{M}
\rangle \negthinspace \rangle$ and $P_0$ in the ``isothermal''
runs is exactly the same as the $\langle \negthinspace \langle P
\rangle \negthinspace \rangle$-$\langle \negthinspace \langle w_{M}
\rangle \negthinspace \rangle$ relation in the adiabatic
calculations. 
This fact indicates that the pressure-stress relation is robust and
unrelated to the time-dependent behavior of the shearing box.

It is worth noticing that the horizontal axis of Figure~\ref{fig:wmgpbz}
is $\gamma P$.
In terms of the comparison with incompressible MHD turbulence, the
dependence on $\gamma$ has an important meaning.
The dependence on the gas pressure is very weak and the difference in
$\gamma$ between the adiabatic ($\gamma = 5/3$) and ``isothermal''
($\gamma = 1.001$) models is quite small.
Therefore it is difficult to distinguish the dependence on $P$ and
$\gamma P$.
To see the difference more clearly, we consider rather extreme cases
with $\gamma = 5$ (S52g and S62g), which are shown by blue symbols in
the figure.
Except for $\gamma$, their model parameters are identical to S52 and
S62, respectively.
For each model, three time averages over successive 50-orbit
periods are shown.
Although the range in the gas pressure is not so wide, a net increase
of the saturation level with time can be seen in both models.
If $\gamma P$ is used as the horizontal axis, the time-averaged stress
in S52g and S62g is closer to the fitting function of the $\gamma =
5/3$ runs ({\it dotted line}).
To verify the $\gamma$ dependence, however,
calculations with larger $\gamma$ are needed.
As expected from equation (\ref{eqn:pt}) the increase in the gas
pressure is faster if $\gamma$ is larger.
Thus calculations with large $\gamma$ are difficult to perform using a
time-explicit numerical scheme, because the time step ($\propto \Delta
/ c_s$) becomes small.
Although the saturation level of the stress may depend on $\gamma P$,
hereafter in this paper we use the gas pressure as the horizontal axis
in similar diagrams.
Direct comparison with incompressible MHD turbulence driven by the
MRI may be interesting, but is beyond the scope of this paper.

\subsubsection{Constraint on Numerical Resolution}
\label{sec:res}

In this subsection, we examine the numerical resolution needed to
obtain a correct pressure-stress relation.
Because the magnetic energy is proportional to the Maxwell stress
in MHD turbulence, the same dependence on the gas pressure can be
seen in $\langle \negthinspace \langle B_z^2 / 8 \pi \rangle
\negthinspace \rangle$.
The saturation level of the vertical magnetic energy is shown in
Figure~\ref{fig:bzpe}.
The dotted line indicates a fitting function obtained from all the
adiabatic ($\gamma = 5/3$) models shown in Figure~\ref{fig:wmpbz}.
Filled circles in Figure~\ref{fig:bzpe} denote the ``isothermal''
models S51i, S61i, S52i, and S62i from left to right.
The saturation level of models S52i and S62i is on the dotted line.
However, the lower pressure models (S51i and S62i) are located far
below the predicted line.
MHD turbulence in these two models decays in the late stages of
the evolution, and the magnetic energy decreases with time.

Because the density is almost uniform even in the nonlinear regime, the
volume-averaged magnetic energy can be regarded as the
root-mean-square (RMS) of the MRI wavelength,
\begin{equation}
\langle \lambda_{\rm MRI}^2 \rangle^{1/2} = 2 \pi \frac{\langle
  v_{{\rm A}z}^2 \rangle^{1/2}}{\Omega} = \frac{2 \pi}{\Omega}
\left( \frac{\langle B_z^2 \rangle}{4 \pi \rho_0} \right)^{1/2} \;.
\end{equation}
Throughout the paper, the MRI wavelength is calculated from
the vertical component of the magnetic field because the fastest
growing mode of the MRI, the axisymmetric mode, is characterized by
the vertical field strength.
The RMS of the MRI wavelength is shorter when the gas pressure is
lower, if the saturation level of the magnetic energy is proportional
to $P^{1/4}$.
The numerical resolution in terms of the MRI wavelength is then
poorer at lower gas pressures.
The gap between models S61i and S52i suggests that there is a minimum
resolution at which the correct $\langle \negthinspace \langle P
\rangle \negthinspace \rangle$-$\langle \negthinspace \langle w_{M}
\rangle \negthinspace \rangle$ relation is obtained.

The horizontal dot-dashed line in Figure~\ref{fig:bzpe} indicates
where the RMS of the MRI wavelength is 6 grid zones.
Therefore, the MRI wavelength must be resolved by at least 6 grid zones
to avoid the decay of MHD turbulence due to numerical diffusion and to
obtain the predicted saturation level. 
If $\langle \negthinspace \langle \lambda_{\rm MRI}^2 \rangle
\negthinspace \rangle^{1/2} \lesssim 6 \Delta$,
the characteristics of the turbulence are quite different from those
in well-resolved models.
For example, the Reynolds stress in models S51i and S61i is larger
than the Maxwell stress, while the Maxwell stress dominates the
Reynolds stress in all the other models (see Table 1).
This can also be seen when the Ohmic dissipation is effective
(Fleming, Stone, \& Hawley 2000; Sano \& Stone 2002b).
The condition on numerical resolution may be useful not only for
the local shearing box simulations but also for global disk
simulations.

Figure~\ref{fig:bzpe} includes results obtained using a different
numerical scheme. 
Open squares are from models S51e, S61e, S52e, and S62e, which are
solved using an internal energy scheme that does not conserve total
energy.  The thermal energy increases more slowly due to energy losses
from the system.
For example, the gas pressure in models S52 and S52e at 300 orbits is
$\langle \negthinspace \langle P \rangle \negthinspace \rangle / P_0 =
30$ and 3, respectively.
However, the $\langle \negthinspace \langle P \rangle \negthinspace
\rangle$-$\langle \negthinspace \langle w_{M} \rangle \negthinspace
\rangle$ relation is found to be unchanged, i.e., independent of the
type of numerical scheme.
The saturation level of the stress in the models solved by the
internal energy scheme is always slightly smaller than the results of
the total energy scheme (see Table 1).
But the difference can be explained clearly by the difference in gas
pressure, if the power-law relation is taken into account. 

Further evidence of the resolution limit can be seen in the time
evolution of model S51e.
Because this model is solved by the non-conservative scheme, the
increase of the gas pressure is slow.
Crosses in the figure mark the time averages for model S51e that are
taken over every 50 orbits after 50 orbits.
When the gas pressure is low and the predicted saturation level is
below the resolution limit, the magnetic energy is much lower than the
dotted line.
However, as the gas pressure increases and the predicted level exceeds
the limit, the time-averaged magnetic energy starts to follow the
predicted dotted line.
This behavior indicates that we need at least about 6 grid zones per
MRI wavelength to resolve MHD turbulence driven by the MRI.

\subsection{Uniform $B_z$ Models}

The effects of the field geometry are examined in this section.  The
magnetic field structures in accretion disks are difficult to observe,
and remain poorly known.  Therefore it is important to survey
different possible field geometries and understand the effects on the
nonlinear evolution of the MRI.
If the disk is penetrated by a dipole field of the central object, or
by a global field of the surrounding interstellar medium, there may be
a net vertical flux.
Here we consider models beginning with a uniform vertical field. 
The vertical flux is conserved in the shearing box, so that a seed
field for the MRI is always present if it is present initially.
Previous numerical work (Hawley et al. 1995; 1996; Sano \& Stone
2002b) has shown that the existence of a net vertical flux greatly
affects the nonlinear evolution of the MRI.
For example, the saturation level of the stress depends on the
strength of the uniform field, although the quantitative relation is
not confirmed yet.  The stress is reported to be proportional to $B_z$
(Hawley et al. 1995) and to $B_z^2$ (Sano, Inutsuka, \& Miyama 1998;
Turner et al. 2003).
Because the saturation amplitude may depend on several physical
quantities simultaneously, it is necessary to understand all the
effects in order to extract the separate contributions of the
field strength and gas pressure.
Therefore, we scrutinize the dependence of the
saturation amplitude on the physical quantities as well as the effects
of the field geometry.

\subsubsection{Gas Pressure Dependence}

First, there is a weak power-law relation between the gas pressure and
the Maxwell stress for uniform $B_z$ models. 
The relation is very similar to that seen in the zero net flux models.
In Figure~\ref{fig:wmpz}, the saturation level of the Maxwell stress
is shown as a function of the gas pressure.
Colors denote the initial field strength which is, from weaker to
stronger,
$v_{{\rm A}0} = 1.5625 \times 10^{-6}$ ($black$), 
$6.25 \times 10^{-6}$ ($cyan$), 
$1.25 \times 10^{-5}$ ($blue$), 
$2.5 \times 10^{-5}$ ($green$), 
$5 \times 10^{-5}$ ($red$), and
$1 \times 10^{-4}$ ($pink$).
The circles are from the adiabatic ($\gamma = 5/3$) models and the
squares are from the ``isothermal'' ($\gamma = 1.001$) models. 
Cases with the same initial field strength may be compared to extract
the gas pressure dependence alone.
Look at the $v_{{\rm A}0} = 2.5 \times 10^{-5}$ models ($green$), for
instance.
The parameters of these 5 models are identical except for the initial
gas pressure. 
The saturation level of the stress shows a weak dependence on $P$.
The power-law index is about $1/6$, and is smaller than that in the
zero net flux runs.

For purposes of comparison, the pressure-stress relation of
the zero net flux models is shown by the solid line in the figure.
All the results from the uniform $B_z$ models lie
above the solid line, that is, the saturation level is always higher
than in the zero net flux models.
A power-law relation can be seen for the other $v_{{\rm A}0}$ models
as well.
The power-law index is slightly smaller than, or comparable to, that
of the zero net flux models.

Figure~\ref{fig:wmpz} includes the ``isothermal'' cases.
In general, the saturation level of the stress in the ``isothermal''
models is a few times smaller than in the adiabatic counterparts.
Because of the large stress, the increase of the gas pressure is
dramatic in the adiabatic models [see eq.~(\ref{eqn:pt})].
For model Z51, for example, the gas pressure at the end of the
calculation is 3 orders of magnitude higher than the initial value.
The gas pressure in the nonlinear regime is many times larger than in
the isothermal model (Z51i).
Therefore the Maxwell stress with respect to $\langle \negthinspace
\langle P \rangle \negthinspace \rangle$ (i.e., the $\alpha$
parameter) has a huge difference between the adiabatic and
``isothermal'' runs; $\alpha \approx 2.8 \times 10^{-3}$ in Z51 while
$\alpha \approx 0.67$ in Z51i.
For the uniform $B_z$ cases, the $\alpha$ parameter ranges widely from
$10^{-4}$ to 1 (see Table 2).
The magnitude of the plasma beta is approximately given by the inverse
of the $\alpha$ parameter.
The plasma beta values in the nonlinear regime span a wide range, from
10 to $10^5$, and there is no characteristic amplitude.

\subsubsection{Dependence on the Initial Field Strength}

As is expected, the saturation level of the stress is larger for
stronger initial fields.
Since the stress has a dependence on the gas pressure as well as the
field strength, both effects should be taken into account at the
same time.  From the results shown in Figure~\ref{fig:wmpz}, the
saturation level of the Maxwell stress is approximately given by 
$\langle \negthinspace \langle w_{M} \rangle \negthinspace \rangle 
\propto 
\langle \negthinspace \langle P \rangle \negthinspace \rangle^{1/6} 
v_{{\rm A}0}^{3/2}$ 
using the gas pressure in the nonlinear regime $\langle \negthinspace
\langle P \rangle \negthinspace \rangle$ and the initial field
strength $v_{{\rm A}0}$.
This is plotted in the figure by dotted lines for models Z4* ($blue$),
Z5* ($green$), and Z6* ($red$).
Here, Z4*, Z5*, and Z6* denote all the models with a uniform field of
$v_{{\rm A}0} = 1.25 \times 10^{-5}$, $2.5 \times 10^{-5}$, and $5
\times 10^{-5}$, respectively.
For example, Z6* includes Z61, Z62, Z63, Z61i, and Z62i.
The relation shown by each dotted line is valid only for one value of
the initial field strength.
We find that there is an upper and a lower limit to the saturation
level for the uniform $B_z$ runs.  The precise dependence on the
field strength is difficult to measure because of the small
range between the limits.
The origins of the limits are discussed in the next subsection.

We perform an additional numerical experiment to demonstrate the
importance of the amount of net magnetic flux.
So far the initial field is assumed to be uniform everywhere in the
computational domain.  However, 
model Z62p has an initially uniform field that is localized within a
small part of the domain.
At the beginning, the magnetic field is confined to $-0.5 < x <
0.5$ and $-1 < y < 1$, and fills one quarter of the volume of the box.
The field strength and the other parameters are exactly the same as
model Z62.
The time evolution of the magnetic energy density is illustrated in
Figure~\ref{fig:p24} by a slice in the $x$-$z$ plane at $y = 0$. 
The radial pattern at 2 orbits (Fig.~\ref{fig:p24}, {\it top-right
panel}) is formed due to the background shear motion, but the field
is still confined within $-0.5 < x < 0.5$.
Then the magnetized region starts to spread in the radial direction
as a result of the MRI.
The most unstable wavelength is 0.31 in this case, so that the
characteristic wavelength at 3 orbits ({\it middle-right panel}) is
consistent with the prediction of the linear analysis.
When the generated horizontal field reaches nonlinear amplitude, the
linear growth of the MRI is disrupted by magnetic reconnection.
Through diffusion effects, the mass fraction of the magnetized region
increases. 
Finally, at 3.5 orbits ({\it bottom-left panel}) the whole domain is
filled with amplified magnetic field and becomes turbulent.  The
turbulence is sustained throughout the calculation.

The saturation level of the Maxwell stress in model Z62p is indicated
by the red cross in Figure~\ref{fig:wmpz}.
The initial field strength in this model is the same as in models Z6*
($red$), while the total magnetic flux is equal to that of models Z4*
($blue$).  The stress in Z62p is much smaller than in models Z6*, and
comparable to models Z4*. 
Thus the key quantity of the uniform vertical field is its total flux,
or average strength over the entire system.
This simple numerical experiment yields two interesting results.
First, if even a small part of the disk has a magnetic field, the
field can spread due to the shear motion of the disk and
the growth of the MRI.
Second, the saturation amplitude is determined by the total vertical
flux.

\subsubsection{Upper and Lower Limit of the Saturation Level}

In general, saturation levels are higher for larger vertical magnetic
fluxes. 
However this relation has an upper and a lower limit.
Figure~\ref{fig:btz} shows the time evolution of the magnetic energy
for models Z32 ($v_{{\rm A}0} = 6.25 \times 10^{-6}$), Z62
($v_{{\rm A}0} = 5 \times 10^{-5}$), and Z92 ($v_{{\rm A}0} = 4 \times
10^{-4}$).
The initial magnetic field in model Z32 is 4 times weaker than in
Z62, and 16 times weaker than in Z92.
The field is amplified by many orders of magnitude in the
lower $B_0$ models (Z32 and Z62).
The magnetic energies in Z62 are greater than in Z32, both initially
and in the nonlinear regime.
On the other hand, no amplification is seen in model Z92. 
The reason for this behavior is explained as follows.

The shortest unstable wavelength of the MRI, or critical wavelength,
is proportional to the Alfv{\'e}n speed.  The critical wavelength is
longer for stronger initial fields.  When the critical wavelength
exceeds the disk scale height, linear growth of the MRI can no longer
be expected.
This gives the upper limit to the field strength.
The critical wavelength in model Z92 is initially longer than the
vertical size of the box: $\lambda_{\rm crit} = \lambda_0 / \sqrt{3}
\approx 1.4 > L_z$.
The model is magnetorotationally stable, because the magnetic
tension suppresses the linear growth of the MRI.
Model Z92 ({\it dotted line}) shows no growth in the magnetic energy. 

The volume-averaged magnetic energy in model Z62 fluctuates greatly
with time.  The amplitude is almost an order of magnitude.
The fluctuations are a typical feature of calculations with a net
vertical magnetic flux.
During the saturated turbulent phase, field amplification due to the
growth of a channel solution occurs quasi-periodically, and is
followed by dissipation through magnetic reconnection (Sano \&
Inutsuka 2001).
In weaker $B_0$ models, on the other hand, the 
two-channel flow appears less often, and the time variations are of
smaller amplitudes. 
If the net vertical flux is less than a critical value, the saturation
level of the stress approaches that of the zero net flux results.

In Figure~\ref{fig:wmpz}, the pressure-stress relation obtained from
the zero net flux runs is shown by a solid line.  The field strength
in models Z3* ($cyan$) is half that in models Z4* ($blue$).  The field
in models Z1* ($black$) is 8 times weaker than in models Z4*.
However the differences in the stress are small, and the weaker
field results are almost on the solid line.
This fact suggests that the saturation amplitude of the zero net flux
runs gives the minimum level.
The recurrent growth of the channel flow enhances the saturation
level of the stress when the disk has a vertical flux.
The enhancement is seen only when the field strength is large
enough that the initial MRI wavelength $\lambda_0$ corresponds to at
least several percent of the system size.

The two-channel flow appears if the vertical field is amplified such
that the MRI wavelength corresponds to the vertical box size.
The vertical field must be amplified by a larger factor to produce the
channel flow if the initial field is weaker.
Thus the appearance of the two-channel flow is rarer in the weaker
initial field models.
Within the region of parameter space we explored, the zero net flux
models and the weak uniform field models show the same level of
saturation in the Maxwell stress.  For those models, the power-law
index in the pressure-stress relation is about 1/4.

If the MRI wavelength of the initial uniform field is longer than
about one tenth of the vertical box size, the appearance of the
two-channel flow enhances the saturation level of the stress.  In this
parameter regime, the saturation amplitude is roughly proportional to
$B_0^{3/2}$, and the power-law index is slightly smaller ($\sim$ 1/6).
The upper limit on the saturation level is given by a condition that
the MRI wavelength corresponding to the uniform vertical field should
be smaller than the vertical domain size.

Even in cases with the most unstable wavelength initially less than
the grid size (e.g., model Z13), the longer-wavelength unstable modes
can be resolved.
Furthermore the characteristic wavelength in the nonlinear stage
is longer because of the amplification of the field.
Thus the resolution condition discussed in \S~\ref{sec:res} is
satisfied in the saturated state in all the models shown in
Figure~\ref{fig:wmpz}.

\subsection{Effects of Magnetic Dissipation}

Dissipation of the magnetic field may play an important role in
determining the saturation level of the MRI, because magnetic
reconnection occurs frequently during the saturated turbulent phase.
The ideal MHD approximation is made in all the calculations listed in
Tables 1 and 2.  Due to numerical diffusion, magnetic reconnection
occurs even in the ideal MHD simulations.
In this section, we briefly examine the effects of physical
dissipation.  The Ohmic dissipation term is explicitly included when
solving the induction equation.  
Two types of resistivity are considered: a time-constant and spatially
uniform value $\eta = \eta_0$, and an anomalous resistivity.
The anomalous diffusivity adopted varies as $\eta = k_0 (v_d
- v_{d0})^2$ for $v_d > v_{d0}$ and $\eta = 0$ for $v_d < v_{d0}$,
where $v_d \equiv |\mbox{\boldmath $J$}| / \rho$ is the drift
velocity, and $k_0$ and $v_{d0}$ are parameters.
This prescription is based on the idea that current-driven
instabilities enhance the effective diffusivity, and has been used in
many astrophysical simulations (e.g., Yokoyama \& Shibata 1994;
Machida \& Matsumoto 2003).

Table 3 lists the resistive models, including both those with zero net
flux and those with uniform vertical initial fields.
The size of the Ohmic dissipation is indicated by the magnetic
Reynolds number $Re_{M} = V L / \eta$, where $V$ and $L$ are typical
velocity and length scales.
For the MRI, typical scales are the Alfv{\'e}n speed $V \sim v_{\rm
A}$ and the most unstable wavelength $L \sim v_{\rm A}/\Omega$.
The magnetic Reynolds number is then $Re_M = v_{\rm A}^2 / \eta
\Omega$.  Linear and nonlinear evolution of the MRI is characterized
very well using the parameter $Re_M$ (Sano \& Miyama 1999; Sano \&
Stone 2002b).

Dissipation effects are better-resolved for larger $\eta$.
However, Ohmic dissipation suppresses the MRI if the diffusivity is
too large.
The critical value of the initial magnetic Reynolds number is about 10
for zero net flux cases (Sano \& Stone 2002b).
Actually, MHD turbulence dies away in 100 orbits when a uniform
diffusivity is added to models S51 and S52 with $\eta_0 = 10^{-7}$,
corresponding to $Re_M = 6.3$.
Therefore we choose $\eta_0 = 10^{-7.5}$ ($Re_M = 20$) for the
uniform diffusivity in zero net flux models.
For uniform $B_z$ cases, on the other hand, the critical $Re_{M}$ is
about unity (Sano \& Stone 2002b).
Thus the diffusivity for models Z51r, Z52r, Z53r, and Z52ir is assumed
to be $10^{-6}$, corresponding to $Re_{M} = 0.63$.  For cases with
anomalous diffusivity, we take $k_0 = 0.05$ and $v_{d0} = 0.05$,
because the mean value of $\eta$ in the nonlinear regime in regions
with $v_d > v_{d0}$ is then a few times $10^{-8}$ in the zero net
flux models, and a few times $10^{-7}$ in the uniform $B_z$ models.

The pressure-stress relations in the resistive runs are shown in
Figure~\ref{fig:eta}$a$ (zero net flux $B_z$) and \ref{fig:eta}$b$
(uniform $B_z$) together with the ideal MHD results ($circles$).
Models S51, S52, S53, and S52i are shown in Figure~\ref{fig:eta}$a$
and models Z51, Z52, Z53, and Z52i are in Figure~\ref{fig:eta}$b$.
Model parameters in the resistive runs are identical to these ideal
MHD models except for the magnetic diffusivity. 
A positive correlation can be seen for both the uniform
($squares$) and anomalous diffusivity runs ($crosses$).
The saturation amplitude in the resistive runs is slightly lower than
that in the $\eta = 0$ cases.
But the difference is at most a factor of 3, because the dependence on
the diffusivity is weak when the magnetic Reynolds number is larger
than unity (Sano \& Stone 2002b).
The dotted lines in the figure indicate the power-law relation
$\langle \negthinspace \langle w_M \rangle \negthinspace \rangle
\propto \langle \negthinspace \langle P \rangle \negthinspace
\rangle^{q}$ with $q = 1/4$ (Fig.~\ref{fig:eta}$a$) and $q = 1/6$
(Fig.~\ref{fig:eta}$b$).
The values of the index $q$ for the resistive runs are similar to
those in the ideal MHD runs.

The diffusion length for the magnetic field is roughly $l_{\rm diff} =
2 \pi \eta / v_{{\rm A}z}$.
In the uniform $B_z$ models, this scale is well-resolved 
initially because $l_{\rm diff} = 2 \pi \eta / v_{{\rm A}0}
\approx 8 \Delta$.
In the nonlinear regime, the average diffusion length is still
larger than the grid scale ($\langle l_{\rm diff} \rangle = 2 \pi \eta
/ \langle v_{{\rm A}z}^2 \rangle^{1/2} \sim 3 \Delta$).
However, because of the severe constraint on the initial $Re_{M}$, the
diffusion length in the zero net flux models is shorter than the grid
scale ($\langle l_{\rm diff} \rangle / \Delta \approx 0.4$).
We therefore carried out double-resolution versions of the
calculations shown in Figure~\ref{fig:eta}a, using $64 \times 256
\times 64$ zones.  The dependence on the gas pressure is qualitatively
unaffected by the change in the numerical resolution.  We conclude that
the weak power-law relation between the gas pressure and the Maxwell
stress exists for both the ideal and resistive MHD cases.

\subsection{General Features of the Turbulence}

\subsubsection{Characteristic Quantities}

The numerical calculations discussed above show that the saturation
level depends upon the gas pressure, the field strength and geometry,
and the equation of state.  At the same time, the saturated states in
all these calculations show common features.

Table 4 lists characteristic quantities in turbulence driven by
the MRI.
The quantities are averages from all the models listed in Tables 1 --
3, excepting poorly-resolved cases (S51i and S61i) and a stable model
(Z92).
The averages include even the ``isothermal'' models, the $\gamma = 5$
models, and the resistive models.  The ratios listed are independent
of the initial conditions and field configuration.  The standard
deviations among the models are included in the table after $\pm$.

The most interesting ratio in table 4 is that of the Maxwell stress to
the magnetic pressure.
The ratio is about $\langle \negthinspace \langle w_M \rangle
\negthinspace \rangle / \langle \negthinspace \langle P_{\rm mag}
\rangle \negthinspace \rangle = 0.46$ and takes a similar value for
all the models.
Because the Maxwell stress is always about five times the Reynolds
stress, the total stress is approximately proportional to the
magnetic pressure in the MRI turbulence.
The shear motion preferentially enhances the toroidal field, so that
$B_y$ is always the dominant component.
The pressures in the components of the field have universal ratios,
$\langle \negthinspace \langle B_y^2 \rangle \negthinspace \rangle /
\langle \negthinspace \langle B_z^2 \rangle \negthinspace \rangle =
23$ and $\langle \negthinspace \langle B_x^2 \rangle \negthinspace
\rangle / \langle \negthinspace \langle B_z^2 \rangle \negthinspace
\rangle = 3.3$.
The magnetic field in the turbulence is anisotropic.
On the other hand, anisotropy in the perturbed velocity is rather
weak, with
$\langle \negthinspace \langle \delta v_y^2 \rangle \negthinspace
\rangle / \langle \negthinspace \langle v_z^2 \rangle \negthinspace
\rangle = 2.2$ and $\langle \negthinspace \langle v_x^2 \rangle
\negthinspace \rangle / \langle \negthinspace \langle v_z^2 \rangle
\negthinspace \rangle = 2.6$.
The ratios listed in Table 4 are valid for well-developed turbulence
driven by the MRI.
If Ohmic dissipation is effective and MRI is suppressed, then the
ratios take quite different values (Sano \& Stone 2002b).

The magnetic energy density in MRI-driven turbulence is correlated
with the perturbed kinetic energy $\delta E_{\rm kin} \equiv \rho
\delta v^2 / 2$, and not with the total kinetic energy.
The ratio is typically $\langle \negthinspace \langle \delta
E_{\rm kin} \rangle \negthinspace \rangle / \langle \negthinspace
\langle E_{\rm mag} \rangle \negthinspace \rangle = 0.33$.
On the other hand, the ratio of thermal to magnetic energy has no
universal value and ranges from 10 to $10^6$ in our models.
As seen from the pressure-stress relation, the ratio of
magnetic to thermal energy varies approximately as $E_{\rm mag} /
E_{\rm th} \propto P^{-3/4}$ for zero net flux runs.
Thus this ratio depends on the gas pressure in the nonlinear regime.
This appears to be inconsistent with results of global disk
simulations, in which typically $\beta \sim 100$ (e.g.,
Igumenshchev, Narayan, \& Abramowicz 2003).
The difference may be due partly to the local approximation.
In the shearing-box simulations, the box height is assumed fixed,
while in the global simulations, the thickness of the disk can change
with the pressure.
For a complete comparison with the global simulations, a quantitative
understanding of the effects of both box size and gas pressure 
is required.  Box size effects will be discussed in a subsequent paper.

\subsubsection{Fluctuations}

MHD turbulence in our simulations is driven by the MRI.
Fluctuations begin mainly by the growth of unstable modes of
the MRI, and initially take the form of perturbations in the magnetic
field and velocity.
The fluctuations in magnetic pressure may affect the distribution
of the gas pressure.
We find that the amplitudes of the fluctuations in the magnetic and
gas pressures are always comparable in the turbulent regime.
The spatial dispersion of the pressure, $\langle \delta P^2 \rangle^{1/2}
\equiv \langle (P - \langle P \rangle )^2 \rangle^{1/2}$, is evaluated
from a snapshot of the spatial distribution of the pressure.
Figure~\ref{fig:dp} shows the dispersions of both the magnetic and
gas pressure for all the models listed in Tables 1 -- 3.
Here the poor resolution models (S51i and S61i) and stable model (Z92)
are excluded.
For each model, a snapshot is chosen from near the end of the
calculation.  The time variation in the saturation level is quite
large for the uniform $B_z$ runs, because two phases occur.  In one
phase, a two-channel flow dominates, and in the other, the field is
weaker and the turbulence is disorganized.
For some of the uniform $B_z$ models (Z51, Z61, Z51i, Z52i, Z61i, and
Z62i), the spatial dispersions of ten randomly selected snapshots from
the saturated turbulence are plotted in the figure, so as to include
information about both of the phases.
Open and filled symbols are from the ``isothermal'' and adiabatic
runs, respectively.  Circles denote zero net flux $B_z$ runs, and
squares are from the uniform $B_z$ runs.
Evidently the relation $\langle \delta P^2 \rangle^{1/2} \approx
\langle P_{\rm mag}^2 \rangle^{1/2}$ holds for all the models shown
in this figure.

The spatial dispersions of the density ($circles$) and magnetic
pressure ($squares$) are shown in Figure~\ref{fig:rhopm} as functions
of the ratio of volume-averaged magnetic pressure to volume-averaged
gas pressure, $\langle P_{\rm mag} \rangle / \langle P \rangle$.
Again, open symbols are from ``isothermal'' models, and filled symbols 
are from adiabatic models.
The gas pressure in the adiabatic models increases significantly
in the nonlinear regime, so that gas pressure is much greater than
magnetic pressure.
The ratio $\langle P_{\rm mag} \rangle / \langle P \rangle$ in the
adiabatic models ({\it filled symbols}) is always less than 0.01.
In the ``isothermal'' models, on the other hand, the magnetic pressure
can be comparable to the gas pressure.

For both the adiabatic and ``isothermal'' cases, the fluctuation in
$\langle P_{\rm mag} \rangle$ is independent of $\langle P_{\rm mag}
\rangle / \langle P \rangle$ and near unity, if the magnetic pressure
is lower than the gas pressure.
If $\langle P_{\rm mag} \rangle \sim \langle P \rangle$, then the
fluctuation is slightly less.
Taking account of the compressibility, the linear growth rate of the
axisymmetric MRI is reduced when the toroidal component of the
magnetic pressure is comparable to the gas pressure (Blaes \& Balbus
1994; Kim \& Ostriker 2000).
This may reduce the fluctuations in the magnetic field.
The density fluctuation, on the other hand, is proportional to the
ratio of magnetic to gas pressure.
In the adiabatic cases the density is almost spatially uniform
({\it filled symbols}), because the magnetic pressure is much less
than the gas pressure in the saturated state.
Only when $\langle P_{\rm mag} \rangle$ is comparable to $\langle P
\rangle$ do order unity density fluctuations occur.

All the data plotted in Figure~\ref{fig:rhopm} are well-fitted by
functions $\langle \delta P_{\rm mag}^2 \rangle^{1/2} / \langle P_{\rm
mag} \rangle \approx \left( \langle P_{\rm mag} \rangle / \langle P
\rangle + 1 \right)^{-1}$
and
$\langle \delta \rho^2 \rangle^{1/2}/\langle \rho \rangle \approx 
\left( \langle P \rangle / \langle P_{\rm mag} \rangle + 1
\right)^{-1}$, which are shown by dotted curves.
Crosses in the figure are obtained from 10 snapshots of
model Z62i.
Although the density fluctuation and magnetic pressure are
time-dependent, the variations move the model along the same relations
in this diagram.
The amplitude of density fluctuations in MRI turbulence is found to
be given by 
\begin{equation}
\frac{\langle \delta \rho^2 \rangle^{1/2}}{\langle \rho \rangle}
\approx 
\frac{\langle \delta P_{\rm mag}^2 \rangle^{1/2}}{\langle P \rangle}
\;.
\label{eqn:df}
\end{equation}
This relation is valid for all the models we performed.
Large density fluctuations are found only when the fluctuations in
magnetic pressure are comparable to the gas pressure. 

Where radiation pressure is important, the relationship between the
fluctuations is more complicated.
In radiation-dominated disks, the presence of large density
fluctuations in MRI turbulence requires fast radiative diffusion as
well as magnetic pressures comparable to the gas pressure (Turner,
Stone, \& Sano 2002; Turner et al. 2003).
In such disks, the role of the gas pressure is sometimes played by
gas and radiation pressures together, depending on the radiative
diffusion timescale.
When the diffusion length and MRI wavelength are comparable, the
effective pressure is intermediate between the gas and total
pressures.
Equation~(\ref{eqn:df}) may hold in radiation-dominated disks if the
gas pressure is replaced by the effective pressure.  The equation
might be used to estimate the effective pressure from the
fluctuations.

The energy density of the fluctuations in MHD turbulence is an
interesting quantity in terms of the energy balance.
Equipartition is found to hold between the kinetic and magnetic
energies of fluctuations (Fig.~\ref{fig:dek}).
The symbols in the figure are the same as in Figure~\ref{fig:dp}.
Here the perturbed magnetic energy is defined as $\langle \delta B^2 /
8 \pi \rangle \equiv \langle ( |B| - \langle |B| \rangle )^2 / 8 \pi
\rangle$.
The disturbances in kinetic and magnetic energies shown in
Figure~\ref{fig:dek} both originate from unstable modes of the MRI,
and the two are roughly equal throughout the saturated turbulence.
The thermal energy of the perturbations, on the other hand, is not in
equipartition, and is comparable to or less than the magnetic energy.
Figure~\ref{fig:det} shows the thermal energy of the disturbances as a
function of the magnetic energy.  The size of the thermal disturbances
is estimated using the internal energy of sound waves $\langle c_s^2
\delta \rho^2 / 2 \rho \rangle$ (e.g., Landau \& Lifshitz 1959).
The gas pressure increases with time in the adiabatic shearing box
calculations, and the ratio $\langle P_{\rm mag} \rangle / \langle P
\rangle$ decreases.
The density fluctuation is proportional to this ratio, so that $\langle
\delta \rho^2 \rangle$ also decreases with time.
Therefore the thermal energy of perturbations cannot reach an
equilibrium state, while the magnetic and kinetic energies are in
equipartition and almost saturated.

\section{DISCUSSION}

\subsection{Time Variability}

The stresses generated by the MRI fluctuate greatly over time.  Since
the stresses control the loss of gravitational energy from accreting
material, the radiation emitted locally by the disk may also vary.
Here we focus on the characteristics of the time variability of the
stress in our numerical simulations.

\subsubsection{Amplitude of Time Variation}

Time variability is shown in Figure~\ref{fig:dispbzall}.  The temporal
dispersion of the stress during the last 50 orbits of each calculation
is normalized by the time-averaged stress, and plotted as a function
of the mean vertical component of the magnetic energy over the same
interval.
The zero net flux runs listed in Table 1 are marked by open circles.
The amplitudes in these models are typically 0.2 -- 0.4, and the
average of 0.34 is shown by a horizontal solid line.  The uniform
$B_z$ models are plotted by colored symbols, with meanings as in
Figure~\ref{fig:wmpz}.
When the magnetic energy in the uniform $B_z$ models is relatively
low, the time variations are comparable to those in the zero net flux
models.
However, time variations are larger in models with higher magnetic
energies.  

The large variations in the uniform vertical field cases
result from recurrent growth of the two-channel flow.  This flow
arises from a linearly unstable MRI mode, and is an exact solution of
the full nonlinear MHD equations in the incompressible limit (Goodman
\& Xu 1994).  The flows grow to nonlinear amplitudes before being
disrupted by shear instabilities and magnetic reconnection.
Two-channel flows occur when the MRI wavelength is comparable to the
vertical height of the shearing box.  The time variability is
sensitive to the ratio of the MRI wavelength $\lambda_{\rm MRI}$ to
the vertical box size $L_z$.  The vertical dotted line in the figure
indicates where the RMS of the MRI wavelength, $\langle \negthinspace
\langle \lambda_{\rm MRI}^2 \rangle \negthinspace \rangle^{1/2} = 2
\pi \langle \negthinspace \langle v_{{\rm A}z}^2 \rangle \negthinspace
\rangle^{1/2} / \Omega$, equals the box height.
Time variations of order unity occur in models with
$\langle \negthinspace \langle \lambda_{\rm MRI}^2
\rangle \negthinspace \rangle^{1/2} \sim L_z$.
When the MRI wavelength is shorter, the growth of unstable modes and
the dissipation by reconnection occur in multiple small regions
simultaneously, and the overall time variations are smoother.
The characteristics of the time variation illustrated in
Figure~\ref{fig:dispbzall} are little changed by the inclusion of
magnetic diffusivity and by increases in numerical resolution up to
$128 \times 512 \times 128$.

The observed amplitudes of X-ray variability in black-hole candidates
and active galactic nuclei are typically about 0.3 (e.g., Nowak et
al. 1999; Papadakis et al. 2002).
Local stress variations of similar size may occur if the MRI
wavelength is an order of magnitude shorter than the disk thickness.
However, radiative processes must be taken into account for more
detailed comparison with the observations.  Moreover the amplitude may
be sensitive to global effects, as it depends on the size of the box.
It may be interesting to compare observed variability with results
from future global disk simulations.

\subsubsection{Temporal Power Spectra}

The power spectrum is a tool used to make comparisons with
observations.  Power spectral density (PSD) may be calculated from the
history of the volume-averaged stress.  Kawaguchi et al. (2000) found
in global disk MHD simulations that the PSD of the stress has a shape
similar to the PSD of the X-ray flux in observations of black-hole
candidates.

Figure~\ref{fig:npsd} shows the PSD for three models Z62r (strong
uniform field), Z32r (weak uniform field), and S62r (zero
net flux).
They are calculated from the history of the Maxwell stress measured at
intervals $10^{-3} t_{\rm rot}$.
The normalized PSD shown in Figure~\ref{fig:npsd} are obtained by the
following procedure. 
The history data from 50 to 300 orbits is divided into 8 intervals of
about 30 orbits each.
The PSD is calculated for each segment and the amplitude is normalized
by the squared time-average of the stress of each term.
Then the PSD averaged over these 8 spectra is shown in the figure. 
The error bars are standard errors $\sigma / N^{1/2}$ of the 8
segments.
Note that the bottom two models in the figure are shifted downward to
avoid overlapping; the amplitude of the spectrum is multiplied by
$10^{-2}$ and $10^{-4}$ for model Z32r and S62r, respectively.

A uniform diffusivity $\eta_0 = 10^{-6}$ is used in model Z62r.
The diffusion length is $l_{\rm diff} / \Delta \approx 4$, and the
effects of Ohmic dissipation may be adequately resolved.
MRI turbulence is prevented by diffusion if the magnetic Reynolds
number, defined using the MRI wavelength and the Alfv{\'e}n speed, is
less than unity (Sano \& Stone 2002b).
In model Z32r, a diffusivity $\eta_0=10^{-6}$ corresponds to magnetic
Reynolds number 0.039.  For sustained turbulence in model Z32r, we
therefore use a lower diffusivity $\eta_0 = 10^{-7.5}$, corresponding
to $Re_{M} = 1.2$ and $l_{\rm diff} / \Delta \approx 1$.

The PSD of the weak uniform field (Z32r) and zero
net flux (S62r) models are quite similar, as are the amplitudes of
the time variability (Fig.~\ref{fig:dispbzall}).
The strong uniform field model (Z62r) exhibits large time variations
with quasi-periodic spike-shaped excursions.
The PSD of this model has a steeper slope near the orbital frequency
than the other models.
The power-law indexes $p$ (PSD $\propto f^{-p}$) of the spectra are
listed in Table 5.
Because the spectra are not well fit by single power-laws, the
indexes are calculated for three different frequency ranges.
The indexes for the ideal MHD and anomalous diffusivity models are
also listed for purposes of comparison.
In all the models, the PSD is steepest in the middle frequency range
$0.1 < 2 \pi f / \Omega < 1$, and shallowest in the lower frequency
range $0.01 < 2 \pi f / \Omega < 0.1$.

Basically, the time variation of the stress in the MRI turbulence
consists of repeated exponential growth and exponential decay.
The PSD of a simple exponential function is flat at low frequencies
and proportional to $f^{-2}$ at high frequencies.
This feature can be seen at the higher and lower frequencies in the
spectra shown in Figure~\ref{fig:npsd}.
The power-law indexes in the higher range $1 < 2 \pi f / \Omega <
10$ are close to 2, and the variation is flattest in the lower range
$0.01 < 2 \pi f / \Omega < 0.1$.
However the slope is much steeper than $p = 2$ in the middle range
$0.1 < 2 \pi f / \Omega < 1$.
This is because the exponential growth and decay are truncated after
finite intervals.
The typical period of the spike-shaped variation is a few orbits, and
the PSD is enhanced near the corresponding frequency. 
Armitage \& Reynolds (2003) find similar features in the temporal
power spectrum of a local annulus in a global disk simulation.
X-ray observations of black-hole candidates and active galactic nuclei
exhibit a power-law decline with a shallower index $p \sim 1$ -- 2
(e.g., Nowak et al. 1999; Papadakis et al. 2002).  This suggests again
that global disk simulations may be necessary for comparison with
observed time variability.

\subsubsection{Growth and Decay Rates of the Stress in Turbulence}

The time variations in stress mostly consist of brief periods of
exponential growth and decay.
The rates of growth and decay of the Maxwell stress can be estimated
approximately by the following method.
From the history of the volume-averaged stress $w (t_i)$ sampled at
regular intervals, we select local extrema $w_{{\rm ex}}$, where $[w
(t_{i+1}) - w (t_i)]/[w (t_{i}) - w (t_{i-1})] < 0$.  The extrema
occur at times $t_{{\rm ex}}$.  The rate of change between the
$j$-th extremum and the next is calculated by
\begin{equation}
\omega = \frac{\ln w_{{\rm ex},j+1} - \ln w_{{\rm ex},j}}
{t_{{\rm ex},j+1} - t_{{\rm ex},j}} \;,
\label{eqn:omega}
\end{equation}
where a positive (negative) sign for $\omega$ indicates growth
(decay).  When the time between peaks is short, the change in stress
is small.  We ignore intervals less than $0.1 t_{\rm rot}$.

Histograms of the growth and decay rates are shown in
Figure~\ref{fig:omegahist} for models Z62r, Z32r, and S62r.
The weak uniform field model (Z32r) and zero net flux model
(S62r) have quite similar distributions, symmetric between growth
and decay.  By contrast the distribution for model Z62r is obviously
asymmetric.
As mentioned above, turbulence in models Z32r and S62r is
disorganized, and the growth and decay of perturbations in different
parts of the simulation volume are uncorrelated.
When averaged over the many perturbations present, the stress varies
less with time than in each individual perturbation.  The histogram
peaks at growth rates less than the maximum for the MRI,
$\omega_{\max} = 0.75 \Omega$.  In model Z62r, on the other hand,
increases are due to the growth of the domain-filling two-channel
flow, and decay happens by nearly simultaneous magnetic reconnection
throughout.
The mean growth rate $0.23 \Omega$ is not much less than the MRI
growth rate.  The distribution of decay rates shows a tail extending
to faster values, suggesting reconnection rates may differ from one
event to another.

The timescale of magnetic reconnection depends on the Alfv{\'e}n
speed. 
For example, the timescale of the slow reconnection model (Sweet
1958; Parker 1957; 1963) is given by 
\begin{equation}
\tau_{\rm rec} = \frac{L}{v_{\rm A}} \sqrt{S}
= \frac{L^{3/2}}{v_{\rm A}^{1/2} \eta^{1/2}} \;,
\label{eqn:trec}
\end{equation}
where $L$ is the size of a reconnection region and $S = L v_{\rm A} /
\eta$. 
If $\omega_{\rm decay} \sim \tau_{\rm rec}^{-1}$, then decay
will be faster when the Alfv{\'e}n speed is faster.
Figure~\ref{fig:omegawm} shows the growth ($circles$) and decay rates
($squares$) for model Z62r as a function of the Maxwell stress.
Time intervals with larger jumps in the stress, $|\ln w_{{\rm
ex},j+1} - \ln w_{{\rm ex},j}| > 1$, are selected for this figure.
The magnetic energy is proportional to the Maxwell stress, so that the
horizontal axis also indicates the amplitude of the magnetic energy.
The growth rate is independent of the magnetic stress, while the decay
rate increases with magnetic stress and energy.  Similar relations
occur in the anomalous diffusivity (Z62a) and ideal MHD models (Z62).
The Alfv{\'e}n speed is $v_{\rm A} = ( 2 w_M / f \rho_0 )^{1/2}$,
where $f \equiv w_M / P_{\rm mag} \approx 0.5$ is the ratio of the
Maxwell stress to the magnetic pressure (Table 4).
If the field decays due to reconnection at a rate $\omega_{\rm rec}
\equiv \tau_{\rm rec}^{-1} = L^{-3/2} v_{\rm A}^{1/2} \eta^{1/2}$,
then
\begin{equation}
\frac{\omega_{\rm rec}}{\Omega} 
= \frac{v_{\rm A}^{1/2} \eta^{1/2}}{L^{3/2} \Omega} 
\approx 
\left( \frac{f}{0.5} \right)^{-1/4}
\left( \frac{L}{0.2} \right)^{-3/2}
\left( \frac{w_M}{P_0} \right)^{1/4}
\label{eqn:rec}
\end{equation}
for model Z62r ($P_0 = 1.25 \times 10^{-5}$, $\eta = 10^{-6}$, and
$\Omega = 10^{-3}$).
The relation given by equation (\ref{eqn:rec}) is shown by a
dot-dashed line in the figure.
Of course, this interpretation may be too simplified to compare with
the simulation results, because the Sweet-Parker picture is based on a
steady structure in the reconnection region.
In MRI-driven turbulence, reconnection continues at most a few orbits,
because the supply of magnetic flux entering the diffusion region is
limited.  Reconnection in this situation may be unsteady.
It is evident that magnetic reconnection is a key process
in determining the saturation level of the MRI turbulence.
Magnetic reconnection also makes a major contribution to magnetic
energy release in global disk simulations (Machida \& Matsumoto 2003).
Understanding non-steady magnetic reconnection may be important for
future progress.

\subsection{Origin of the Pressure Dependence}

The magnetic energy in the saturated turbulence is determined by a
balance between field enhancement by the MRI and dissipation through
magnetic reconnection.
The linear growth rate of the MRI is independent of the gas pressure
if the magnetic pressure is much smaller than the gas pressure, as it
is in the turbulence.
Thus it is quite natural to expect that the gas pressure dependence of
the stress may arise from a pressure dependence in the rate of
dissipation by magnetic reconnection.
The effects of gas pressure on magnetic reconnection are examined by
Ugai \& Kondoh (2001), using two-dimensional MHD simulations with an
anomalous resistivity.  Conditions differ from those we considered in
that magnetic pressure exceeds gas pressure.  Smaller gas pressures
lead to thinner current sheets and more drastic reconnection.
The gas pressure also affects reconnection in laboratory experiments
by Ji et al. (1998).  High gas pressures downstream from the diffusion
region substantially reduce the outflow and thus the reconnection
rate.  Overall, the results of these simulations and experiments are
consistent with a picture in which higher gas pressures lead to slower
reconnection of magnetic fields.

The {\em fluctuation-dissipation relation}, equation (\ref{eqn:edot}),
links the rate of change of thermal energy to the stress in the
nonlinear regime.
If the limiting process in the energy change is reconnection, the
left hand side of equation (\ref{eqn:edot}) is approximately $d E_{\rm
th} / d t \sim E_{\rm th} / \tau_{\rm rec}$, and the saturated
stress is proportional to $P / \tau_{\rm rec}$.
The stress may vary overall with $P^{1/4}$, as in the zero net flux
$B_z$ models, if the reconnection timescale is proportional to
$P^{3/4}$.  For example, if the size of the reconnection region is
proportional to the pressure scale height ($L \propto c_s / \Omega$),
then $\tau_{\rm rec} \propto L^{3/2} \propto P^{3/4}$.
The interpretation discussed here is qualitatively consistent with our
numerical results.  For further quantitative discussion, it may be
necessary to understand the effects of gas pressure on non-steady
magnetic reconnection.

\section{SUMMARY}

We investigate the saturation level of the MRI using
three-dimensional MHD simulations.
To simplify the problem, the local shearing box approximation is
adopted and the vertical component of gravity is ignored.
The dependence of the saturation level on the physical quantities is
scrutinized, with special attention to effects of gas pressure.
The gas pressure increases with time in adiabatic
calculations, and the increase affects the long term
evolution of the saturation level. 
This feature is carefully taken into account in the analysis of the
gas pressure dependence.
The main results are summarized below.

\begin{enumerate}
\item
A power-law relation between the saturation amplitude of the Maxwell
stress and the gas pressure in the nonlinear regime is derived from all
the models performed in this paper.
The power-law index is small ($q = 1/4$ -- 1/6) and varies slightly
with the geometry of the magnetic field.
\item
For the zero net flux models, the power-law index is about $1/4$ and
the saturation level is independent of the initial gas pressure,
the magnetic field strength, and the equation of state.
However, if the MRI wavelength is not resolved by at least six grid
zones, the saturation level is affected greatly by numerical
dissipation.
\item
For the models with a uniform vertical initial field, the power-law
index in the pressure-stress relation is smaller than that in the zero
net flux models.
The saturation level is higher for larger vertical magnetic fluxes,
but there are upper and lower limits on the saturation level depending
on the strength of the uniform field.
\item
Similar pressure-stress relations are obtained in ideal MHD
calculations and resistive MHD models with magnetic Reynolds numbers
greater than about 10.  There is no clear difference in saturation
level between models with a uniform diffusivity and an anomalous
diffusivity.
\item
There exist many characteristic ratios among the quantities in the
MRI turbulence.
The ratios are independent of initial conditions, including the
strength and geometry of the magnetic field, and the gas pressure.
The perturbed magnetic and kinetic energies are maintained near
equipartition.
\item
The amplitude of time variability in the Maxwell stress is
characterized by the ratio of the magnetic pressure to the gas
pressure in the nonlinear regime.
The power spectral density of the temporal variability generally has a
steep slope around the frequency corresponding to a few orbits.
\end{enumerate}

The gas pressure dependence of the saturation level may originate in
the process of magnetic reconnection.  The gain side of the magnetic
energy balance is unlikely to depend on gas pressure, since the MRI
linear growth rate is independent of the pressure when plasma beta
exceeds unity.  Although there is some qualitative evidence supporting
an idea that higher gas pressures reduce reconnection rates, a deeper
understanding of magnetic reconnection is necessary for a quantitative
discussion of the saturation mechanism of the MRI.

The final goal of this work is to derive a predictor function for the
saturation level of the MRI.
For this purpose, we must determine how the saturation
level depends on all the physical quantities. 
For example, we have studied the gas pressure dependence using the
local shearing box approximation, with the height of the box
independent of the gas pressure.
In real accretion disks, however, the scale height of the disk varies
according to the pressure.
Before proceeding to derive a saturation predictor and to examine the
similarities and differences between the $\alpha$ prescription and
numerical results, the effects of changes in disk thickness must be
considered.
The effects of the other parameters, and a final form of the predictor
function, will be discussed in a subsequent paper.

\acknowledgements
We thank Gordon Ogilvie for useful discussions and comments.
Computations were carried out on VPP300/16R and VPP5000 at the National 
Astronomical Observatory of Japan and VPP700 at the Subaru Telescope, 
NAOJ.

\clearpage 

\clearpage

\begin{deluxetable}{cccccccccllll}
\tablewidth{0pc}
\footnotesize
\rotate
\tablecaption{Zero Net Flux $B_z$ Simulations}
\tablehead{
\colhead{Model} & 
\colhead{$\beta_0$} & \colhead{$P_0$} & \colhead{$v_{{\rm A}0}$} & 
\colhead{$H_0$\tablenotemark{a}} & \colhead{$\lambda_0$\tablenotemark{b}} &
\colhead{Size} & \colhead{$\Delta$} & 
\colhead{Orbits} &
\colhead{$\langle\negthinspace\langle w_M 
\rangle\negthinspace\rangle/P_0$} & 
\colhead{$\langle\negthinspace\langle w_R
\rangle\negthinspace\rangle/P_0$} & 
\colhead{$\langle\negthinspace\langle P
\rangle\negthinspace\rangle/P_0$} & 
\colhead{$\alpha\tablenotemark{c}~[\times 10^{3}]$} \\
\colhead{(1)} & \colhead{(2)} & \colhead{(3)} & \colhead{(4)} & 
\colhead{(5)} & \colhead{(6)} & \colhead{(7)} & \colhead{(8)} & 
\colhead{(9)} & \colhead{(10)} & \colhead{(11)} & \colhead{(12)} & 
\colhead{(13)}  
}
\startdata
S41 & $10^2$ & $7.8125 \times 10^{-9}$ & $1.25 \times 10^{-5}$ & 
0.125 & 0.079 &
$1 \times 4 \times 1$ & 1/32 & 100 &
1.31 & 0.305 & 387 & 4.18 \\
S42 & $10^4$ & $7.8125 \times 10^{-7}$ & $1.25 \times 10^{-5}$ & 
1.25 & 0.079 &
$1 \times 4 \times 1$ & 1/32 & 100 &
0.0123 & 0.00272 & 5.94 & 2.53 \\
S43 & $10^6$ & $7.8125 \times 10^{-5}$ & $1.25 \times 10^{-5}$ & 
12.5 & 0.079 &
$1 \times 4 \times 1$ & 1/32 & 100 &
$3.74 \times 10^{-4}$ & $7.02 \times 10^{-5}$ & 1.19 & 0.373 \\
S51 & $10^2$ & $3.125 \times 10^{-8}$ & $2.5 \times 10^{-5}$ & 
0.25 & 0.16 &
$1 \times 4 \times 1$ & 1/32 & 300 &
0.611 & 0.121 & 917 & 0.798 \\
S52 & $10^4$ & $3.125 \times 10^{-6}$ & $2.5 \times 10^{-5}$ & 
2.5 & 0.16 &
$1 \times 4 \times 1$ & 1/32 & 600 &
0.00799 & 0.00159 & 29.9 & 0.320 \\
S53 & $10^6$ & $3.125 \times 10^{-4}$ & $2.5 \times 10^{-5}$ & 
25 & 0.16 &
$1 \times 4 \times 1$ & 1/32 & 300 &
$1.21 \times 10^{-4}$ & $2.21 \times 10^{-5}$ & 1.28 & 0.112 \\
S61 & $10^2$ & $1.25 \times 10^{-7}$ & $5 \times 10^{-5}$ & 
0.5 & 0.31 &
$1 \times 4 \times 1$ & 1/32 & 300 &
0.158 & 0.0316 & 259 & 0.731 \\
S62 & $10^4$ & $1.25 \times 10^{-5}$ & $5 \times 10^{-5}$ & 
5 & 0.31 &
$1 \times 4 \times 1$ & 1/32 & 300 &
0.00165 & $3.26 \times 10^{-4}$ & 4.30 & 0.460 \\
S63 & $10^6$ & $1.25 \times 10^{-3}$ & $5 \times 10^{-5}$ & 
50 & 0.31 &
$1 \times 4 \times 1$ & 1/32 & 300 &
$4.21 \times 10^{-5}$ & $6.67 \times 10^{-6}$ & 1.10 & 0.0442 \\
S71 & $10^2$ & $5 \times 10^{-7}$ & $1.0 \times 10^{-4}$ & 
1 & 0.63 &
$1 \times 4 \times 1$ & 1/32 & 100 &
0.0271 & 0.00569 & 14.4 & 2.27 \\
S72 & $10^4$ & $5 \times 10^{-5}$ & $1.0 \times 10^{-4}$ & 
10 & 0.63 &
$1 \times 4 \times 1$ & 1/32 & 100 &
$3.93 \times 10^{-4}$ & $7.89 \times 10^{-5}$ & 1.23 & 0.382 \\ 
\multicolumn{13}{l}{$\gamma = 1.001$ (``isothermal''):} \\
S51i & $10^2$ & $3.125 \times 10^{-8}$ & $2.5 \times 10^{-5}$ & 
0.25 & 0.16 &
$1 \times 4 \times 1$ & 1/32 & 100 &
0.00135 & $2.71 \times 10^{-4}$ & 1.02 & 1.59 \\
S52i & $10^4$ & $3.125 \times 10^{-6}$ & $2.5 \times 10^{-5}$ & 
2.5 & 0.16 &
$1 \times 4 \times 1$ & 1/32 & 300 &
0.00313 & $7.89 \times 10^{-4}$ & 1.02 & 3.83 \\
S53i & $10^6$ & $3.125 \times 10^{-4}$ & $2.5 \times 10^{-5}$ & 
25 & 0.16 &
$1 \times 4 \times 1$ & 1/32 & 300 &
$1.12 \times 10^{-4}$ & $2.08 \times 10^{-5}$ & 1.01 & 0.132 \\
S61i & $10^2$ & $1.25 \times 10^{-7}$ & $5 \times 10^{-5}$ & 
0.5 & 0.31 &
$1 \times 4 \times 1$ & 1/32 & 100 &
0.00856 & 0.00199 & 1.02 & 10.3 \\
S62i & $10^4$ & $1.25 \times 10^{-5}$ & $5 \times 10^{-5}$ & 
5 & 0.31 &
$1 \times 4 \times 1$ & 1/32 & 300 &
0.00111 & $2.31 \times 10^{-4}$ & 1.02 & 1.32 \\
S63i & $10^6$ & $1.25 \times 10^{-3}$ & $5 \times 10^{-5}$ & 
50 & 0.31 &
$1 \times 4 \times 1$ & 1/32 & 300 &
$4.13 \times 10^{-4}$ & $6.94 \times 10^{-6}$ & 1.00 & 0.0480 \\
\multicolumn{13}{l}{$\gamma = 5$:} \\
S52g & $10^4$ & $3.125 \times 10^{-6}$ & $2.5 \times 10^{-5}$ & 
2.5 & 0.16 &
$1 \times 4 \times 1$ & 1/32 & 200 &
0.0151 & 0.00257 & 85.6 & 0.206 \\
S62g & $10^4$ & $1.25 \times 10^{-5}$ & $5 \times 10^{-5}$ & 
5 & 0.31 &
$1 \times 4 \times 1$ & 1/32 & 200 &
0.00454 & $7.43 \times 10^{-4}$ & 28.1 & 0.188 \\
\multicolumn{13}{l}{Solved by an internal energy (non-conservative)
  scheme:} \\
S51e & $10^2$ & $3.125 \times 10^{-8}$ & $2.5 \times 10^{-5}$ & 
0.25 & 0.16 &
$1 \times 4 \times 1$ & 1/32 & 300 &
0.321 & 0.0727 & 127 & 3.09 \\
S52e & $10^4$ & $3.125 \times 10^{-6}$ & $2.5 \times 10^{-5}$ & 
2.5 & 0.16 &
$1 \times 4 \times 1$ & 1/32 & 300 &
0.00475 & $9.61 \times 10^{-4}$ & 3.24 & 1.76 \\
S61e & $10^2$ & $1.25 \times 10^{-7}$ & $5 \times 10^{-5}$ & 
0.5 & 0.31 &
$1 \times 4 \times 1$ & 1/32 & 300 &
0.0934 & 0.0197 & 45.9 & 2.46 \\
S62e & $10^4$ & $1.25 \times 10^{-5}$ & $5 \times 10^{-5}$ & 
5 & 0.31 &
$1 \times 4 \times 1$ & 1/32 & 300 &
0.00143 & $2.92 \times 10^{-4}$ & 2.44 & 0.706 \\
\enddata
\tablecomments{Time averages are taken over the last 50 orbits.  
For ``isothermal'' runs, time evolutions after 50 orbits are
considered for the time averages.}
\tablenotetext{a}{$H_0 = (2/\gamma)^{1/2} c_{s0} / \Omega = ( 2 P_0 /
  \rho_0 )^{1/2}/ \Omega$}
\tablenotetext{b}{$\lambda_0 = 2 \pi v_{{\rm A}0} / \Omega$}
\tablenotetext{c}{$\alpha = (\langle \negthinspace \langle w_{M} 
\rangle \negthinspace \rangle + \langle \negthinspace \langle w_{R} 
\rangle \negthinspace \rangle) / \langle \negthinspace \langle P
\rangle \negthinspace \rangle$}
\label{tbl:s}
\end{deluxetable}

\begin{deluxetable}{cccccccccllll}
\tablewidth{0pc}
\footnotesize
\rotate
\tablecaption{Uniform $B_z$ Simulations}
\tablehead{
\colhead{Model} & 
\colhead{$\beta_0$} & \colhead{$P_0$} & \colhead{$v_{{\rm A}0}$} & 
\colhead{$H_0$\tablenotemark{a}} & \colhead{$\lambda_0$\tablenotemark{b}} &
\colhead{Size} & \colhead{$\Delta$} & 
\colhead{Orbits} &
\colhead{$\langle\negthinspace\langle w_M 
\rangle\negthinspace\rangle/P_0$} & 
\colhead{$\langle\negthinspace\langle w_R
\rangle\negthinspace\rangle/P_0$} & 
\colhead{$\langle\negthinspace\langle P
\rangle\negthinspace\rangle/P_0$} & 
\colhead{$\alpha\tablenotemark{c}~[\times 10^{3}]$} \\
\colhead{(1)} & \colhead{(2)} & \colhead{(3)} & \colhead{(4)} & 
\colhead{(5)} & \colhead{(6)} & \colhead{(7)} & \colhead{(8)} & 
\colhead{(9)} & \colhead{(10)} & \colhead{(11)} & \colhead{(12)} & 
\colhead{(13)}
}
\startdata
Z13 & $10^6$ & $1.2207
\times 10^{-6}$ & $1.5625 \times 10^{-6}$ & 
1.5625 & 0.0098 &
$1 \times 4 \times 1$ & 1/32 & 200 &
0.0111 & 0.00227 & 9.26 & 1.45 \\
Z14 & $10^8$ & $1.2207
\times 10^{-4}$ & $1.5625 \times 10^{-6}$ & 
15.625 & 0.0098 &
$1 \times 4 \times 1$ & 1/32 & 200 &
$2.11 \times 10^{-4}$ & $4.04 \times 10^{-5}$ & 1.28 & 0.196 \\
Z32 & $10^4$ & $1.9531
\times 10^{-7}$ & $6.25 \times 10^{-6}$ & 
0.625 & 0.039 &
$1 \times 4 \times 1$ & 1/32 & 300 &
0.121 & 0.0241 & 234 & 0.621 \\
Z33 & $10^6$ & $1.9531
\times 10^{-5}$ & $6.25 \times 10^{-6}$ & 
6.25 & 0.039 &
$1 \times 4 \times 1$ & 1/32 & 100 &
0.00109 & $2.18 \times 10^{-4}$ & 1.59 & 0.823 \\
Z42 & $10^4$ & $7.8125 \times 10^{-7}$ & $1.25 \times 10^{-5}$ & 
1.25 & 0.079 &
$1 \times 4 \times 1$ & 1/32 & 100 &
0.0385 & 0.00758 & 22.4 & 2.06 \\
Z43 & $10^6$ & $7.8125 \times 10^{-5}$ & $1.25 \times 10^{-5}$ & 
12.5 & 0.079 &
$1 \times 4 \times 1$ & 1/32 & 100 &
$6.67 \times 10^{-4}$ & $1.19 \times 10^{-4}$ & 1.37 & 0.575 \\
Z51 & $10^2$ & $3.125 \times 10^{-8}$ & $2.5 \times 10^{-5}$ & 
0.25 & 0.16 &
$1 \times 4 \times 1$ & 1/32 & 100 &
3.63 & 0.599 & $1.54 \times 10^3$ & 2.75 \\
Z52 & $10^4$ & $3.125 \times 10^{-6}$ & $2.5 \times 10^{-5}$ & 
2.5 & 0.16 &
$1 \times 4 \times 1$ & 1/32 & 100 &
0.0347 & 0.00573 & 17.7 & 2.29 \\
Z53 & $10^6$ & $3.125 \times 10^{-4}$ & $2.5 \times 10^{-5}$ & 
25 & 0.16 &
$1 \times 4 \times 1$ & 1/32 & 100 &
$4.52 \times 10^{-4}$ & $6.71 \times 10^{-5}$ & 1.29 & 0.402 \\
Z61 & $10^2$ & $1.25 \times 10^{-7}$ & $5 \times 10^{-5}$ & 
0.5 & 0.31 &
$1 \times 4 \times 1$ & 1/32 & 100 &
2.53 & 0.414 & $1.27 \times 10^3$ & 2.31 \\
Z62 & $10^4$ & $1.25 \times 10^{-5}$ & $5 \times 10^{-5}$ & 
5 & 0.31 &
$1 \times 4 \times 1$ & 1/32 & 300 &
0.0333 & 0.00511 & 51.1 & 0.752 \\
Z63 & $10^6$ & $1.25 \times 10^{-3}$ & $5 \times 10^{-5}$ & 
50 & 0.31 &
$1 \times 4 \times 1$ & 1/32 & 100 &
$4.25 \times 10^{-4}$ & $6.75 \times 10^{-5}$ & 1.23 & 0.399 \\
Z72 & $10^4$ & $5 \times 10^{-5}$ & $1 \times 10^{-4}$ & 
10 & 0.63 &
$1 \times 4 \times 1$ & 1/32 & 100 &
0.00890 & 0.00159 & 5.61 & 1.87 \\
Z92 & $10^4$ & $8 \times 10^{-4}$ & $4 \times 10^{-4}$ & 
40 & 2.5 &
$1 \times 4 \times 1$ & 1/32 & 50 &
\nodata & \nodata & \nodata & \nodata \\
\multicolumn{13}{l}{$\gamma = 1.001$ (``isothermal''):} \\
Z32i & $10^4$ & $1.9531
\times 10^{-7}$ & $6.25 \times 10^{-6}$ & 
0.625 & 0.039 &
$1 \times 4 \times 1$ & 1/32 & 100 &
0.0361 & 0.00945 & 1.05 & 43.3 \\
Z42i & $10^4$ & $7.8125 \times 10^{-7}$ & $1.25 \times 10^{-5}$ & 
1.25 & 0.079 &
$1 \times 4 \times 1$ & 1/32 & 100 &
0.0294 & 0.00776 & 1.05 & 35.5 \\
Z51i & $10^2$ & $3.125 \times 10^{-8}$ & $2.5 \times 10^{-5}$ & 
0.25 & 0.16 &
$1 \times 4 \times 1$ & 1/32 & 100 &
0.936 & 0.208 & 1.71 & 668 \\
Z52i & $10^4$ & $3.125 \times 10^{-6}$ & $2.5 \times 10^{-5}$ & 
2.5 & 0.16 &
$1 \times 4 \times 1$ & 1/32 & 100 &
0.0239 & 0.00488 & 1.05 & 27.4 \\
Z61i & $10^2$ & $1.25 \times 10^{-7}$ & $5 \times 10^{-5}$ & 
0.5 & 0.31 &
$1 \times 4 \times 1$ & 1/32 & 100 &
1.06 & 0.257 & 1.98 & 667 \\
Z62i & $10^4$ & $1.25 \times 10^{-5}$ & $5 \times 10^{-5}$ & 
5 & 0.31 &
$1 \times 4 \times 1$ & 1/32 & 100 &
0.0190 & 0.00319 & 1.06 & 21.0 \\
Z72i & $10^4$ & $5 \times 10^{-5}$ & $1 \times 10^{-4}$ & 
10 & 0.63 &
$1 \times 4 \times 1$ & 1/32 & 100 &
0.00767 & 0.00141 & 1.04 & 8.68 \\
\multicolumn{13}{l}{Localized initial field:} \\
Z62p & $10^4$ & $1.25 \times 10^{-5}$ & $5 \times 10^{-5}$ & 
5 & 0.31 &
$1 \times 4 \times 1$ & 1/32 & 100 &
0.00328 & $6.20 \times 10^{-4}$ & 2.53 & 1.55 \\
\enddata
\tablecomments{Time averages are taken over the last 50 orbits.}
\tablenotetext{a}{$H_0 = (2/\gamma)^{1/2} c_{s0} / \Omega = ( 2 P_0 /
  \rho_0)^{1/2} / \Omega$}
\tablenotetext{b}{$\lambda_0 = 2 \pi v_{{\rm A}0} / \Omega$}
\tablenotetext{c}{$\alpha = (\langle \negthinspace \langle w_{M} 
\rangle \negthinspace \rangle + \langle \negthinspace \langle w_{R} 
\rangle \negthinspace \rangle) / \langle \negthinspace \langle P
\rangle \negthinspace \rangle$}
\label{tbl:z}
\end{deluxetable}

\begin{deluxetable}{cccccccccllll}
\tablewidth{0pc}
\footnotesize
\rotate
\tablecaption{Simulations Including the Ohmic Dissipation}
\tablehead{
\colhead{Model} & 
\colhead{$\beta_0$} & \colhead{$P_0$} & \colhead{$v_{{\rm A}0}$} & 
\colhead{$H_0$\tablenotemark{a}} & \colhead{$\eta_0$\tablenotemark{b}} &
\colhead{Size} & \colhead{$\Delta$} & 
\colhead{Orbits} &
\colhead{$\langle\negthinspace\langle w_M 
\rangle\negthinspace\rangle/P_0$} & 
\colhead{$\langle\negthinspace\langle w_R
\rangle\negthinspace\rangle/P_0$} & 
\colhead{$\langle\negthinspace\langle P
\rangle\negthinspace\rangle/P_0$} & 
\colhead{$\alpha\tablenotemark{c}~[\times 10^{3}]$}\\
\colhead{(1)} & \colhead{(2)} & \colhead{(3)} & \colhead{(4)} & 
\colhead{(5)} & \colhead{(6)} & \colhead{(7)} & \colhead{(8)} & 
\colhead{(9)} & \colhead{(10)} & \colhead{(11)} & \colhead{(12)} & 
\colhead{(13)}
}
\startdata
S51r & $10^2$ & $3.125 \times 10^{-8}$ & $2.5 \times 10^{-5}$ & 
0.25 & $10^{-7.5}$ &
$1 \times 4 \times 1$ & 1/32 & 300 &
0.521 & 0.104 & 633 & 0.987 \\
S52r & $10^4$ & $3.125 \times 10^{-6}$ & $2.5 \times 10^{-5}$ & 
2.5 & $10^{-7.5}$ &
$1 \times 4 \times 1$ & 1/32 & 300 &
0.00591 & 0.00119 & 11.2 & 0.635 \\
S53r & $10^6$ & $3.125 \times 10^{-4}$ & $2.5 \times 10^{-5}$ & 
25 & $10^{-7.5}$ &
$1 \times 4 \times 1$ & 1/32 & 150 &
$1.27 \times 10^{-4}$ & $2.27 \times 10^{-5}$ & 1.12 & 0.134 \\
S62r & $10^4$ & $1.25 \times 10^{-5}$ & $5 \times 10^{-5}$ & 
5 & $10^{-7.5}$ &
$1 \times 4 \times 1$ & 1/32 & 300 &
0.00141 & $2.81 \times 10^{-4}$ & 3.98 & 0.424 \\
S51a & $10^2$ & $3.125 \times 10^{-8}$ & $2.5 \times 10^{-5}$ & 
0.25 & anomalous &
$1 \times 4 \times 1$ & 1/32 & 300 &
0.447 & 0.0905 & 624 & 0.862 \\
S52a & $10^4$ & $3.125 \times 10^{-6}$ & $2.5 \times 10^{-5}$ & 
2.5 & anomalous &
$1 \times 4 \times 1$ & 1/32 & 300 &
0.00598 & 0.00121 & 8.65 & 0.832 \\
S53a & $10^6$ & $3.125 \times 10^{-4}$ & $2.5 \times 10^{-5}$ & 
25 & anomalous &
$1 \times 4 \times 1$ & 1/32 & 150 &
$9.60 \times 10^{-5}$ & $1.78 \times 10^{-5}$ & 1.09 & 0.105 \\
S62a & $10^4$ & $1.25 \times 10^{-5}$ & $5 \times 10^{-5}$ & 
5 & anomalous &
$1 \times 4 \times 1$ & 1/32 & 300 &
0.00170 & $3.37 \times 10^{-4}$ & 4.21 & 0.484 \\
Z32r & $10^4$ & $1.9531
\times 10^{-7}$ & $6.25 \times 10^{-6}$ & 
0.625 & $10^{-7.5}$ &
$1 \times 4 \times 1$ & 1/32 & 300 &
0.167 & 0.0321 & 234 & 0.848 \\
Z51r & $10^2$ & $3.125 \times 10^{-8}$ & $2.5 \times 10^{-5}$ & 
0.25 & $10^{-6}$ &
$1 \times 4 \times 1$ & 1/32 & 100 &
1.08 & 0.198 & 529 & 2.42 \\
Z52r & $10^4$ & $3.125 \times 10^{-6}$ & $2.5 \times 10^{-5}$ & 
2.5 & $10^{-6}$ &
$1 \times 4 \times 1$ & 1/32 & 100 &
0.0138 & 0.00254 & 7.34 & 2.22 \\
Z53r & $10^6$ & $3.125 \times 10^{-4}$ & $2.5 \times 10^{-5}$ & 
25 & $10^{-6}$ &
$1 \times 4 \times 1$ & 1/32 & 100 &
$3.01 \times 10^{-4}$ & $4.92 \times 10^{-5}$ & 1.14 & 0.307 \\
Z62r & $10^4$ & $1.25 \times 10^{-5}$ & $5 \times 10^{-5}$ & 
5 & $10^{-6}$ &
$1 \times 4 \times 1$ & 1/32 & 300 &
0.0254 & 0.00428 & 37.7 & 0.788 \\
Z32a & $10^4$ & $1.9531
\times 10^{-7}$ & $6.25 \times 10^{-6}$ & 
0.625 & anomalous &
$1 \times 4 \times 1$ & 1/32 & 300 &
0.0952 & 0.0187 & 170 & 0.669 \\
Z51a & $10^2$ & $3.125 \times 10^{-8}$ & $2.5 \times 10^{-5}$ & 
0.25 & anomalous &
$1 \times 4 \times 1$ & 1/32 & 100 &
1.98 & 0.373 & $1.06 \times 10^3$ & 2.22 \\
Z52a & $10^4$ & $3.125 \times 10^{-6}$ & $2.5 \times 10^{-5}$ & 
2.5 & anomalous &
$1 \times 4 \times 1$ & 1/32 & 100 &
0.0217 & 0.00397 & 13.1 & 1.96 \\
Z53a & $10^6$ & $3.125 \times 10^{-4}$ & $2.5 \times 10^{-5}$ & 
25 & anomalous &
$1 \times 4 \times 1$ & 1/32 & 100 &
$3.86 \times 10^{-4}$ & $6.51 \times 10^{-5}$ & 1.22 & 0.371 \\
Z62a & $10^4$ & $1.25 \times 10^{-5}$ & $5 \times 10^{-5}$ & 
5 & anomalous &
$1 \times 4 \times 1$ & 1/32 & 300 &
0.0197 & 0.00363 & 34.4 & 0.677 \\
\multicolumn{13}{l}{$\gamma = 1.001$ (``isothermal''):} \\
S52ir & $10^4$ & $3.125 \times 10^{-6}$ & $2.5 \times 10^{-5}$ & 
2.5 & $10^{-7.5}$ &
$1 \times 4 \times 1$ & 1/32 & 300 &
0.00230 & $5.98 \times 10^{-4}$ & 1.02 & 2.84 \\
S52ia & $10^4$ & $3.125 \times 10^{-6}$ & $2.5 \times 10^{-5}$ & 
2.5 & anomalous &
$1 \times 4 \times 1$ & 1/32 & 300 &
0.00211 & $5.53 \times 10^{-4}$ & 1.02 & 2.62 \\
Z52ir & $10^4$ & $3.125 \times 10^{-6}$ & $2.5 \times 10^{-5}$ & 
2.5 & $10^{-6}$ &
$1 \times 4 \times 1$ & 1/32 & 100 &
0.0112 & 0.00247 & 1.01 & 13.5 \\
Z52ia & $10^4$ & $3.125 \times 10^{-6}$ & $2.5 \times 10^{-5}$ & 
2.5 & anomalous &
$1 \times 4 \times 1$ & 1/32 & 100 &
0.0195 & 0.00428 & 1.04 & 22.9 \\
\enddata
\tablecomments{Time averages are taken over the last 50 orbits.
For ``isothermal'' runs, time evolutions after 50 orbits are
considered for the time averages.}
\tablenotetext{a}{$H_0 = (2/\gamma)^{1/2} c_{s0} / \Omega = ( 2 P_0 /
  \rho_0 )^{1/2}/ \Omega$}
\tablenotetext{b}{A uniform and constant diffusivity $\eta_0$ or an
 anomalous diffusivity is used.  The anomalous diffusivity
 is assumed to be $\eta = k_0 (v_d - v_{d0})^2$ with $k_0 = 0.05$ and 
 $v_{d0} = 0.05$.}
\tablenotetext{c}{$\alpha = (\langle \negthinspace \langle w_{M} 
\rangle \negthinspace \rangle + \langle \negthinspace \langle w_{R} 
\rangle \negthinspace \rangle) / \langle \negthinspace \langle P
\rangle \negthinspace \rangle$}
\label{tbl:res-s}
\end{deluxetable}

\begin{deluxetable}{cc}
\tablecolumns{2}
\tablewidth{0pc}
\footnotesize
\tablecaption{Characteristic Ratios in MRI Turbulence}
\tablehead{
\colhead{Quantity} & 
\colhead{Average}
}
\startdata
$\langle \negthinspace \langle - B_x B_y / 4 \pi \rangle \negthinspace
\rangle / \langle \negthinspace \langle B^2 / 8 \pi \rangle \negthinspace
\rangle$ &  
$0.467 \pm 0.040$ \\
$\langle \negthinspace \langle - B_x B_y / 4 \pi \rangle \negthinspace
\rangle / \langle \negthinspace \langle \rho v_x \delta v_y \rangle
\negthinspace \rangle$ &  
$5.19 \pm 0.67$ \\
$\langle \negthinspace \langle B_x^2 \rangle \negthinspace
\rangle / \langle \negthinspace \langle B_z^2 \rangle
\negthinspace \rangle$ &  
$3.35 \pm 0.28$ \\
$\langle \negthinspace \langle B_y^2 \rangle \negthinspace
\rangle / \langle \negthinspace \langle B_z^2 \rangle
\negthinspace \rangle$ &  
$23.7 \pm 4.0$ \\
$\langle \negthinspace \langle v_x^2 \rangle \negthinspace
\rangle / \langle \negthinspace \langle v_z^2 \rangle
\negthinspace \rangle$ &  
$2.62 \pm 0.48$ \\
$\langle \negthinspace \langle \delta v_y^2 \rangle \negthinspace
\rangle / \langle \negthinspace \langle v_z^2 \rangle
\negthinspace \rangle$ &  
$2.15 \pm 0.34$ \\
$\langle \negthinspace \langle \delta E_{\rm kin} \rangle \negthinspace
\rangle / \langle \negthinspace \langle E_{\rm mag} \rangle
\negthinspace \rangle$ &  
$0.326 \pm 0.036$ \\
\enddata
\label{tbl:ratio}
\end{deluxetable}

\begin{deluxetable}{cccc}
\tablecolumns{4}
\tablewidth{0pc}
\footnotesize
\tablecaption{Power-Law Indexes of Power Spectral Density}
\tablehead{
\colhead{} & 
\multicolumn{3}{c}{Frequency Range [$2 \pi f / \Omega$]} \\
\cline{2-4} \\
\colhead{Model} & \colhead{0.1 -- 1} & 
\colhead{1 -- 10} & \colhead{10 -- 100} 
}
\startdata
\sidehead{Strong uniform field:}
Z62         & 1.31 & 4.41 & 2.16 \\
Z62r   & 1.00 & 4.14 & 2.35 \\
Z62a & 0.54 & 4.01 & 2.06 \\
\sidehead{Weak uniform field:}
Z32         & 1.80 & 3.66 & 2.07 \\
Z32r   & 1.46 & 3.58 & 2.04 \\
Z32a & 1.66 & 3.66 & 2.05 \\
\sidehead{Zero net flux:}
S62         & 1.83 & 4.12 & 2.11 \\
S62r   & 1.81 & 3.81 & 2.08 \\
S62a & 1.70 & 3.33 & 2.02 \\
\enddata
\label{tbl:power}
\end{deluxetable}

\clearpage

\plottwo{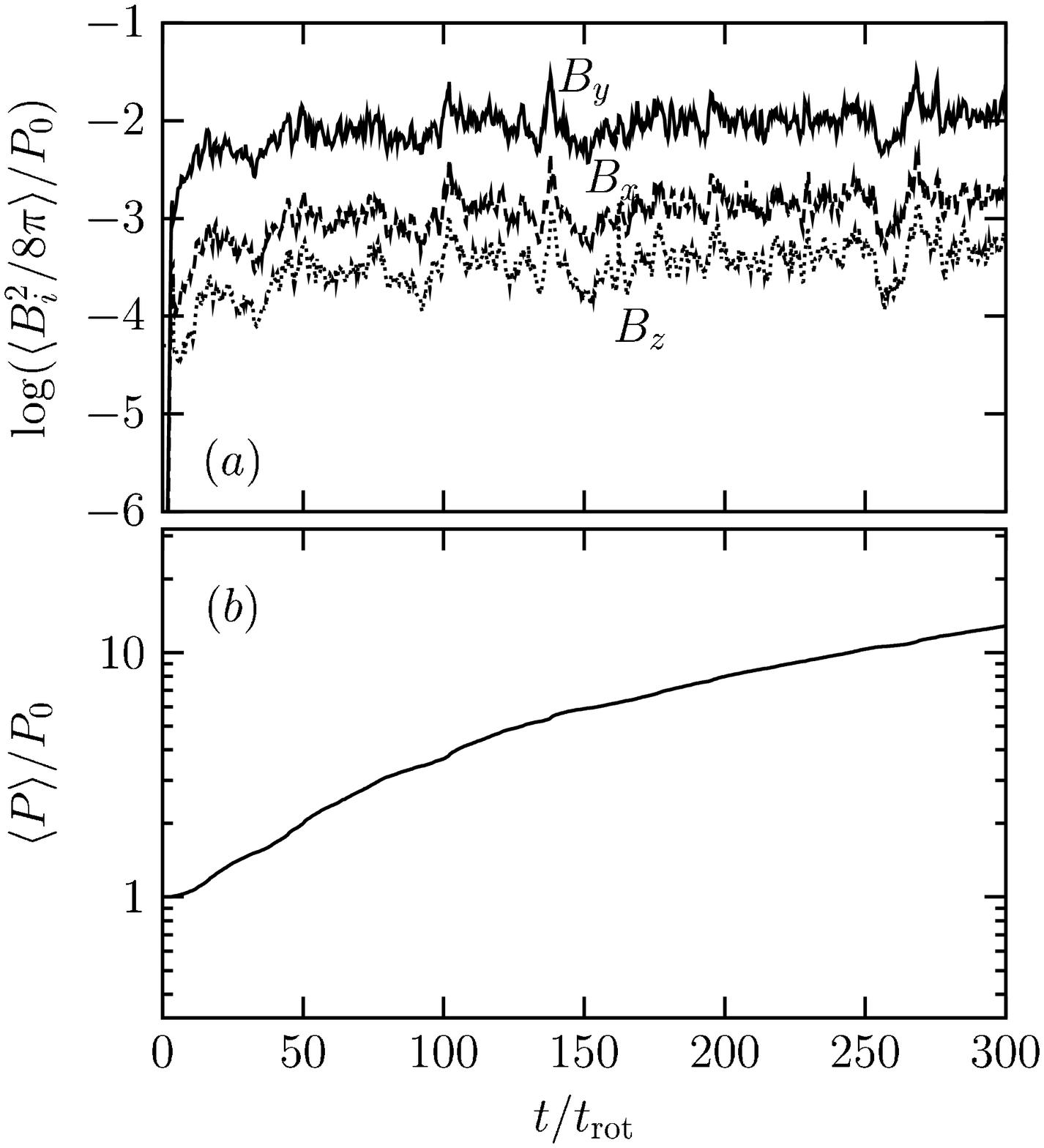}{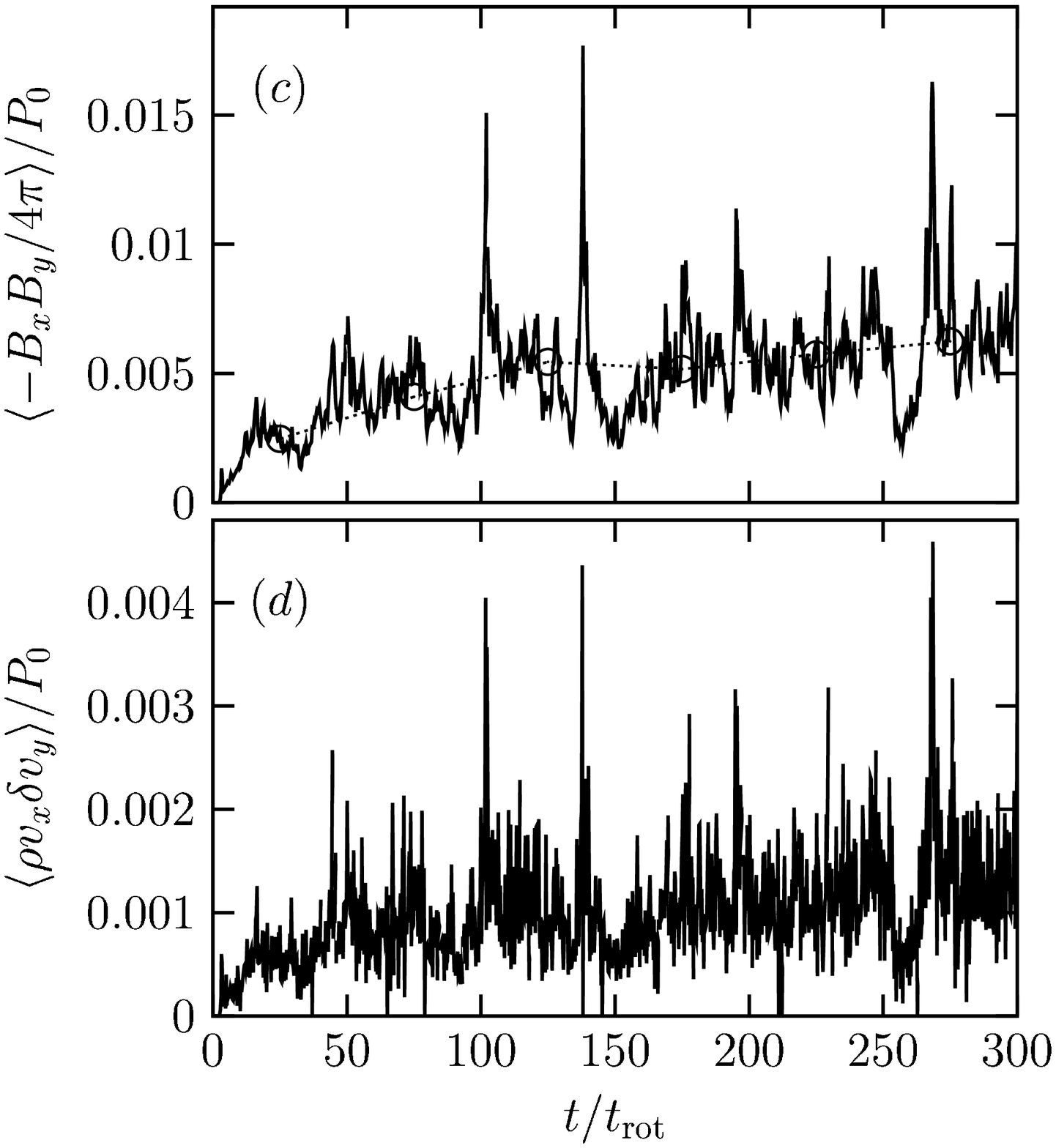}\\
\figcaption
{
Time evolution of the volume-averaged ($a$) magnetic energy $\langle
B_i^2 / 8 \pi \rangle / P_0$, ($b$) gas pressure $\langle P \rangle /
P_0$, ($c$) Maxwell stress $\langle - B_x B_y / 4 \pi \rangle / P_0$,
and ($d$) Reynolds stress $\langle \rho v_x \delta v_y \rangle / P_0$
in the fiducial model S52.
The plasma beta and gas pressure are initially $\beta_0 = 10^4$ and
$P_0 = 3.125 \times 10^{-6}$, respectively.
Time-averages of the Maxwell stress over 50-orbit intervals are
indicated by circles, and show a gradual increase with time.
\label{fig:ts14}}

\clearpage

\plotone{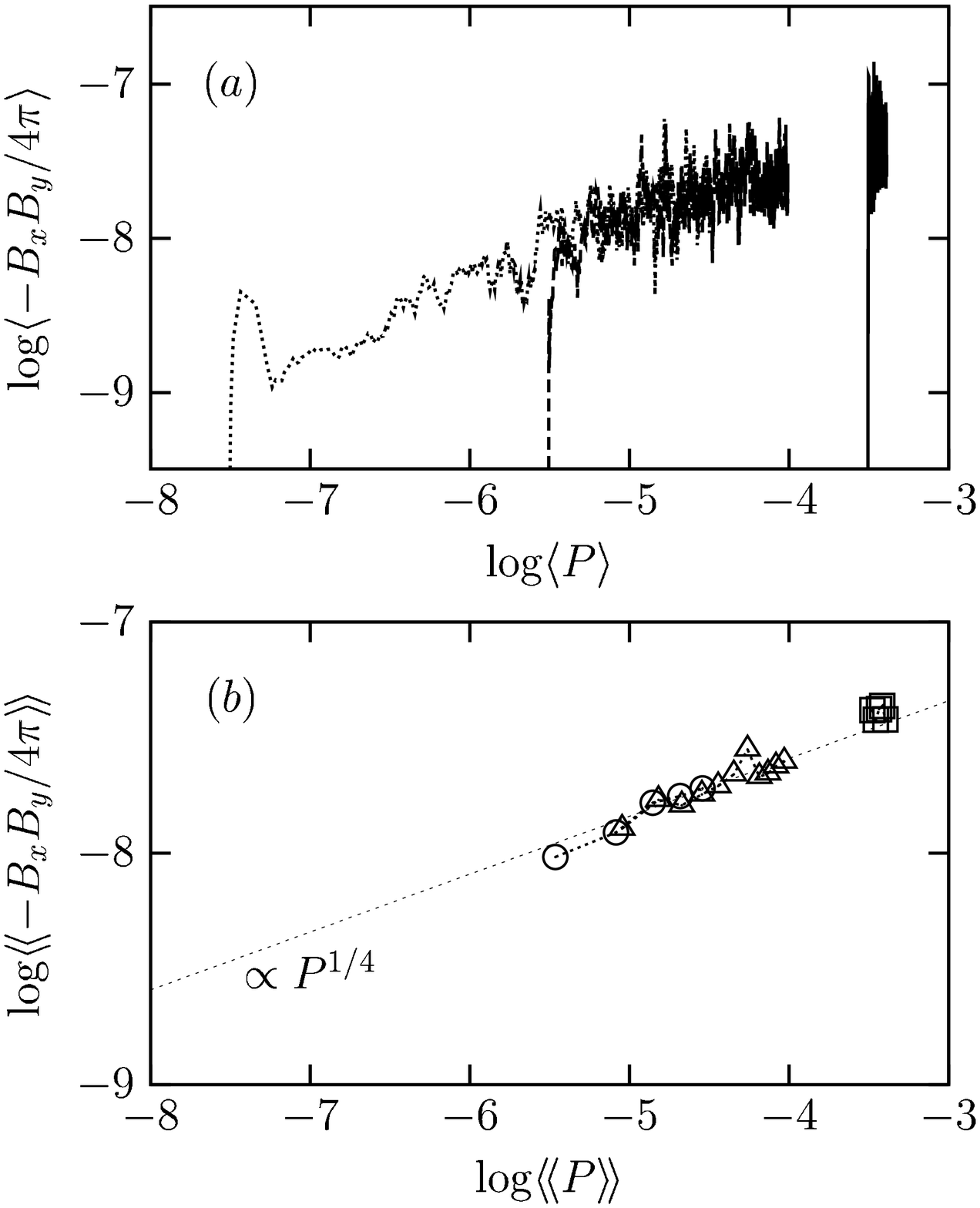}\\
\figcaption
{
($a$) Time evolution of the volume-averaged Maxwell stress $\langle - B_x
B_y / 4 \pi \rangle$ as a function of the volume-averaged gas pressure
$\langle P \rangle$ for models S51 ({\it dotted curve}), S52 ({\it
dashed curve}), and S53 ({\it solid curve}).
The parameters of these models are identical except for the
initial gas pressure.
($b$) Time evolution of the time- and volume-averaged Maxwell stress
$\langle \negthinspace \langle - B_x B_y / 4 \pi \rangle \negthinspace
\rangle$ as a function of the time- and volume-averaged gas pressure
$\langle \negthinspace \langle P \rangle \negthinspace \rangle$ for
models S51 ({\it circles}), S52 ({\it triangles}), and S53 ({\it
  squares}).
The time averages are taken over every 50 orbits after 50 orbits.
The Maxwell stresses obtained are roughly proportional to $P^{1/4}$.
\label{fig:wmps1}}

\clearpage

\plotone{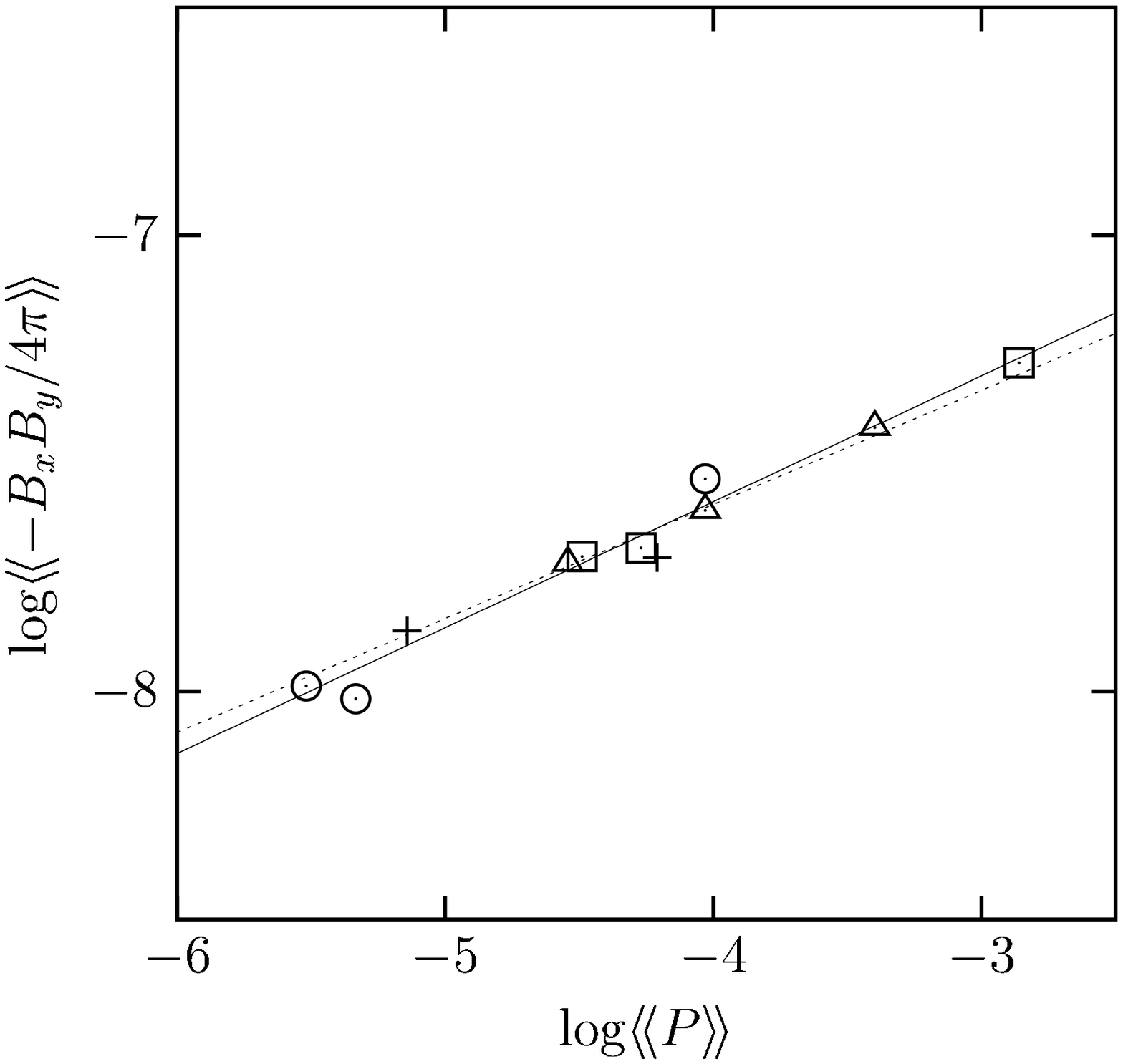}\\
\figcaption
{
Dependence of the time-averaged Maxwell stress $\langle \negthinspace
\langle - B_x B_y / 4 \pi \rangle \negthinspace \rangle$ on the
time-averaged gas pressure $\langle \negthinspace \langle P \rangle
\negthinspace \rangle$ in the nonlinear regime.
Symbols indicate different initial field strengths.  Circles are from
models with initial Alfv{\'e}n speed $v_{{\rm A}0} = 1.25 \times
10^{-5}$ (S41 -- S43), triangles $v_{{\rm A}0} = 2.5 \times
10^{-5}$ (S51 -- S53; fiducial), squares $v_{{\rm A}0} = 5 \times
10^{-5}$ (S61 -- S63), and crosses $v_{{\rm A}0} = 1 \times 10^{-4}$
(S71 and S72).
The time average is taken over the last 50 orbits for each model.
All the results are well-fitted by a power-law relation 
$\langle \negthinspace \langle - B_x B_y / 4 \pi \rangle \negthinspace
\rangle \propto \langle \negthinspace \langle P \rangle
\negthinspace \rangle^{q}$.
The power-law index $q$ is about $1/4$ ({\it dotted line}) and the
best fit is $q = 0.28$ ({\it solid line}).
\label{fig:wmpbz}}

\clearpage

\plotone{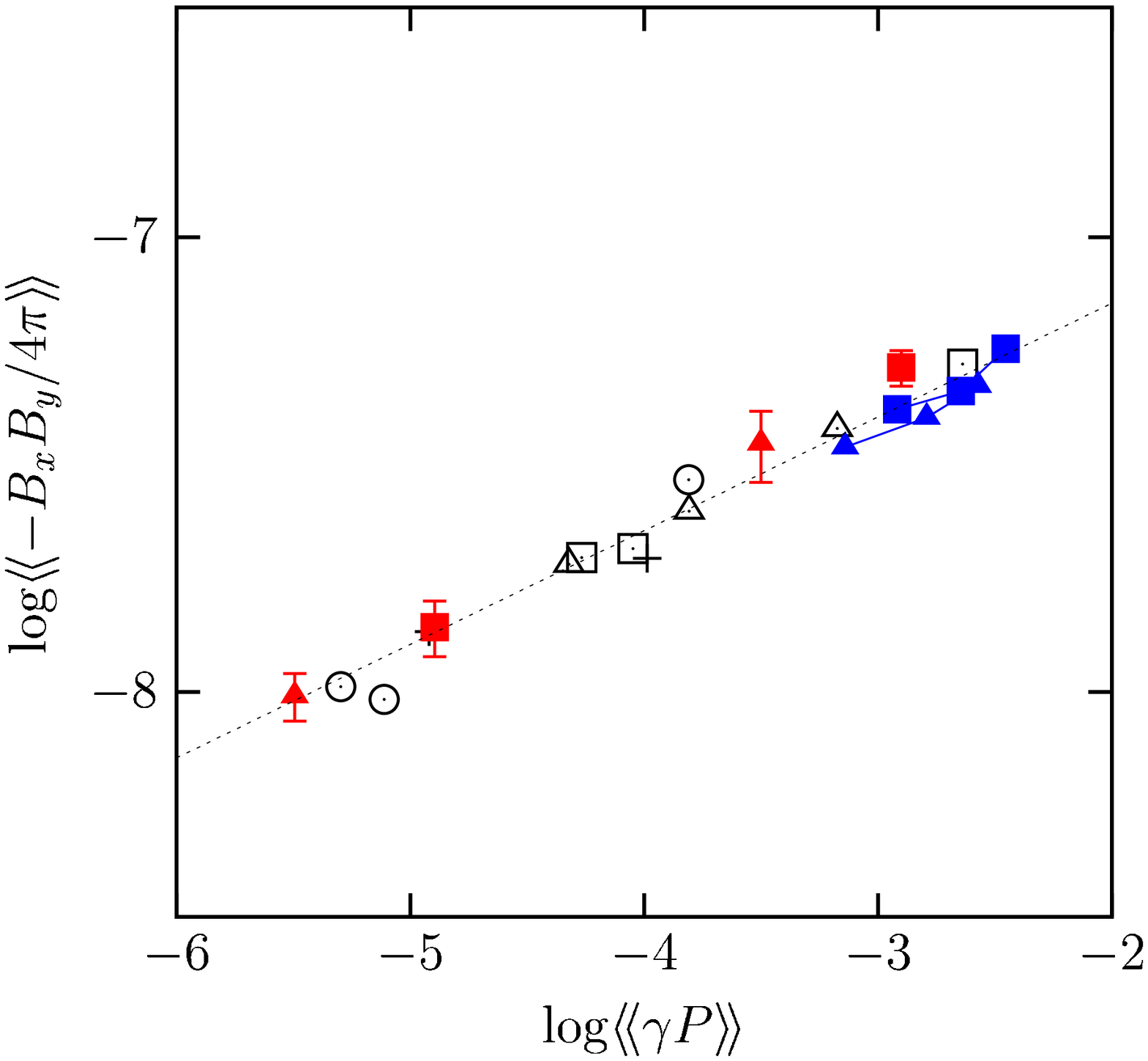}\\
\figcaption
{
Dependence of the time-averaged Maxwell stress $\langle \negthinspace
\langle - B_x B_y / 4 \pi \rangle \negthinspace \rangle$ on $\langle
\negthinspace \langle \gamma P \rangle \negthinspace \rangle$ in the
nonlinear regime.
The ``isothermal'' models with $\gamma = 1.001$ are shown by red
symbols; from left to right, they are models S52i, S62i, S53i, and
S63i.
For each model, the time average is taken from 50 to 300 orbits and
the error bar denotes the dispersion calculated from the time averages
taken every 50 orbits.
Blue symbols are results of the $\gamma = 5$ models; S52g
($triangles$) and S62g ($squares$).
The time average is calculated every 50 orbits after 50 orbits.
The $\gamma = 5/3$ models are shown by black symbols with
meanings as in Figure~{\protect{\ref{fig:wmpbz}}}.
A dotted line shows $\langle \negthinspace \langle w_M \rangle
\negthinspace \rangle \propto \langle \negthinspace \langle \gamma P
\rangle \negthinspace \rangle^{1/4}$ as inferred from the models with
$\gamma = 5/3$.
\label{fig:wmgpbz}}

\clearpage

\plotone{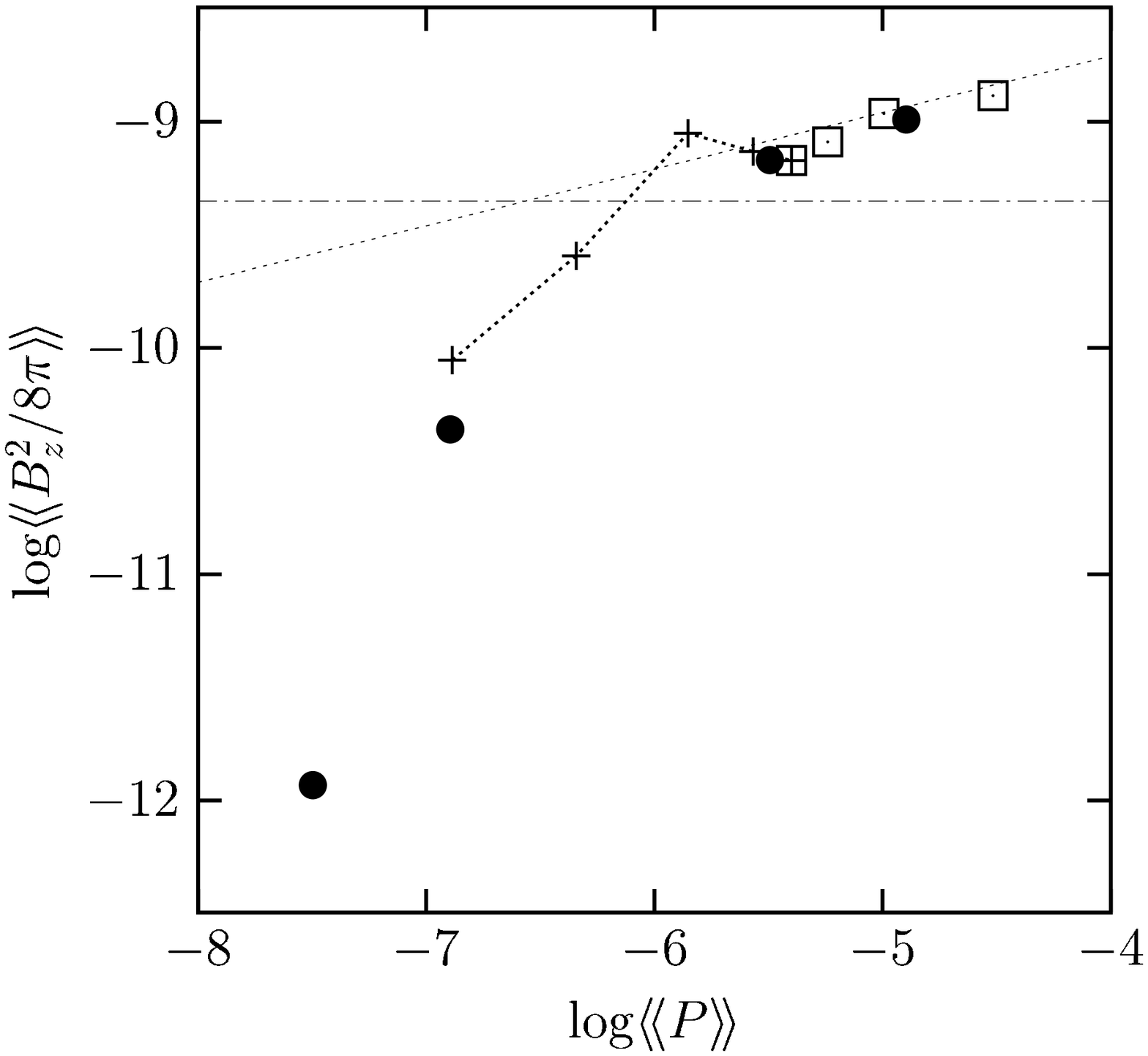}\\
\figcaption
{
Saturation level of the vertical component of the magnetic energy
$\langle \negthinspace \langle B_z^2 / 8 \pi \rangle \negthinspace
\rangle$ as a function of the gas pressure $\langle \negthinspace
\langle P \rangle \negthinspace \rangle$.
Models solved with $\gamma = 1.001$ (``isothermal'') are depicted by
filled circles (from left to right, S51i, S61i, S52i, and S62i).
Open squares mark results using the internal energy scheme (from left
to right, S51e, S61e, S52e, and S62e).
The time average is taken over the last 50 orbits for each model.
Crosses show the time evolution of the time-averaged Maxwell stress in
model S51e.  The time averages extend over 50-orbit periods beginning
at 50 orbits.
A dotted line shows the predicted saturation level obtained from the
models shown in Figure~{\protect{\ref{fig:wmpbz}}}.
On the dot-dashed line, the RMS of the MRI wavelength $\langle
\negthinspace \langle \lambda_{\rm MRI}^2 \rangle
\negthinspace \rangle^{1/2} \equiv 2 \pi ( \langle \negthinspace \langle
B_z^2 \rangle \negthinspace \rangle / 4 \pi \rho_0 )^{1/2} / \Omega$
is equal to 6 grid zones.  Below this limit, saturation levels lie
far below the predicted line.
\label{fig:bzpe}}

\clearpage

\plotone{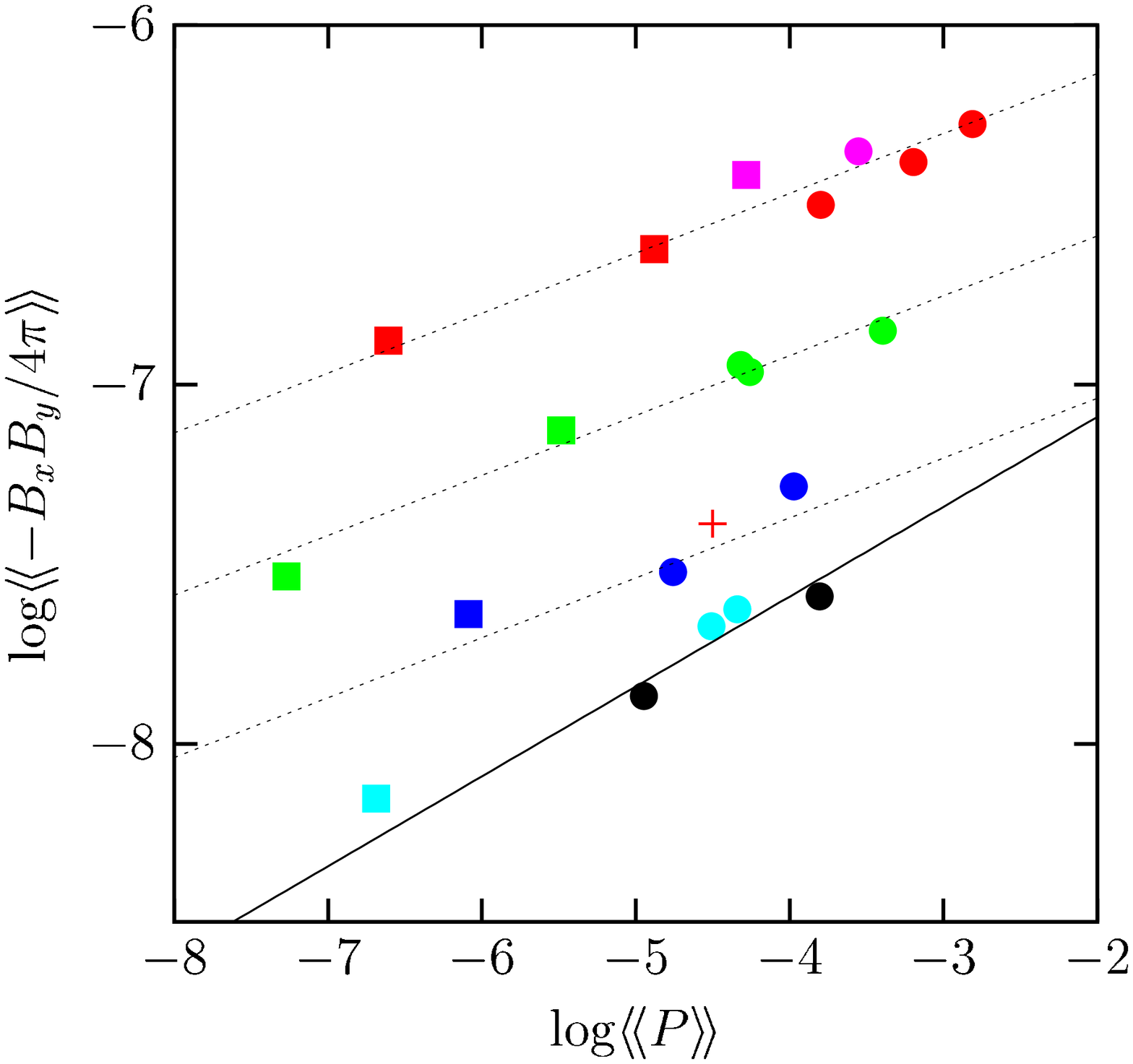}\\
\figcaption
{
Saturation level of the Maxwell stress in the models started with
uniform vertical magnetic fields.  
The colors of the symbols denote the strengths of the
initial fields: $v_{{\rm A}0} = B_0 / (4 \pi \rho_0)^{1/2} = 1.5625
\times 10^{-6}$ ($black$), 
$6.25 \times 10^{-6}$ ($cyan$), 
$1.25 \times 10^{-5}$ ($blue$), 
$2.5 \times 10^{-5}$ ($green$), 
$5 \times 10^{-5}$ ($red$), and
$1 \times 10^{-4}$ ($pink$).
The adiabatic runs are shown by circles and the ``isothermal'' runs by
squares. 
The cross is the result of model Z62p, which is started with a
localized vertical magnetic field in the region $-0.25 < x < 0.25$ and
$-1 < y < 1$.  The total magnetic flux of this model is the same as
Z42 ($blue$), while the field strength is the same as Z62 ($red$).
A solid line shows the pressure-stress relation ($w_{M} \propto
P^{1/4}$) for the zero net flux $B_z$ models, and dotted lines are
fitted functions for models Z4*, Z5*, and Z6*, with $w_{M} \propto
P^{1/6}$.
\label{fig:wmpz}}

\clearpage

\plotone{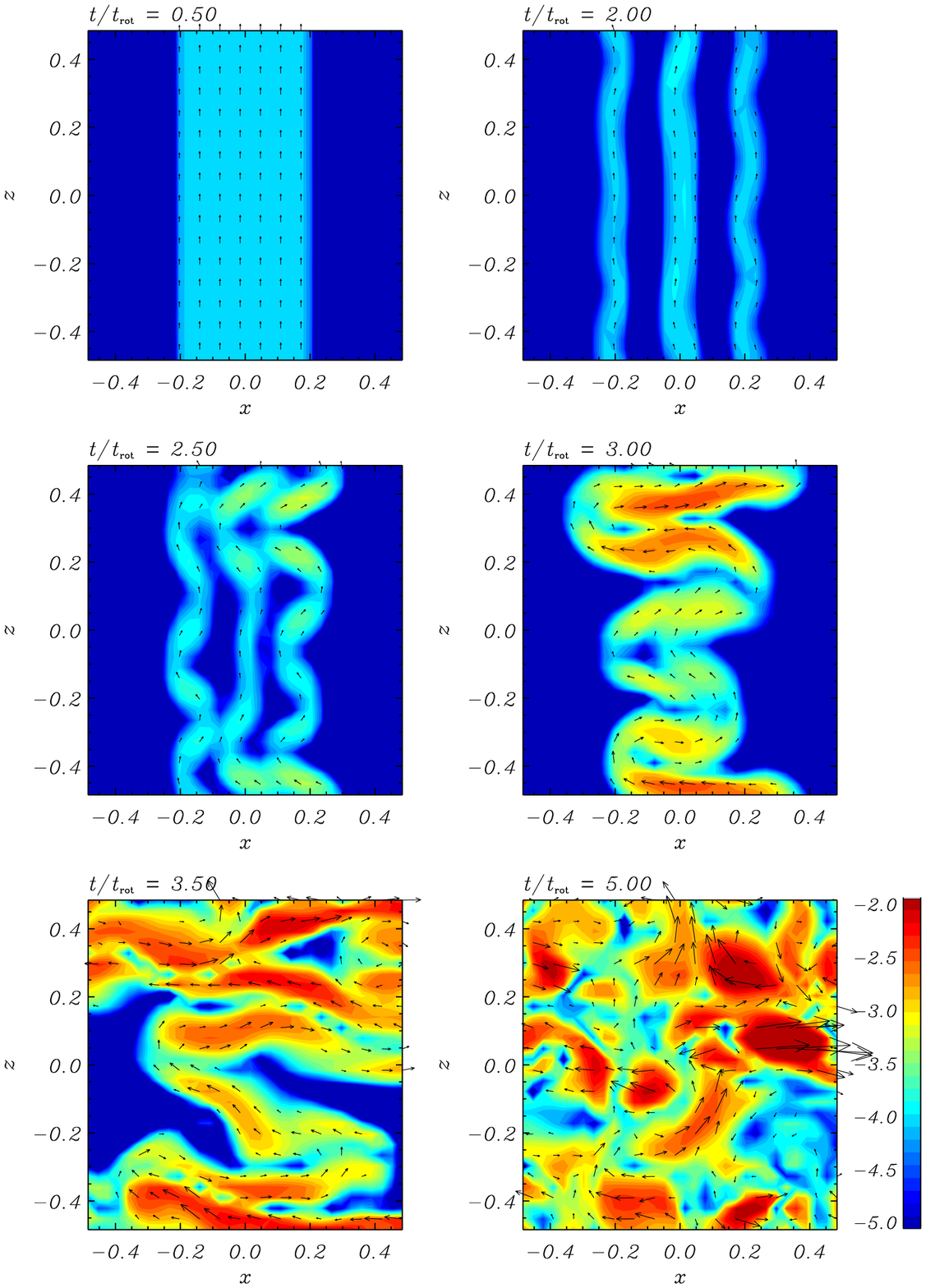}\\
\figcaption[fig7.ps]
{
Magnetic fields in model Z62p, on $x$-$z$ slices at $y = 0$.  The
model is started with a localized vertical field, uniform in $-0.5 < x
< 0.5$ and $-1 < y < 1$.  Colors show the logarithm of magnetic
pressure, arrows the strength and direction of the poloidal magnetic
field.  The MRI enlarges the magnetized region and after a few orbits
the entire domain is turbulent.
\label{fig:p24}}

\clearpage

\plotone{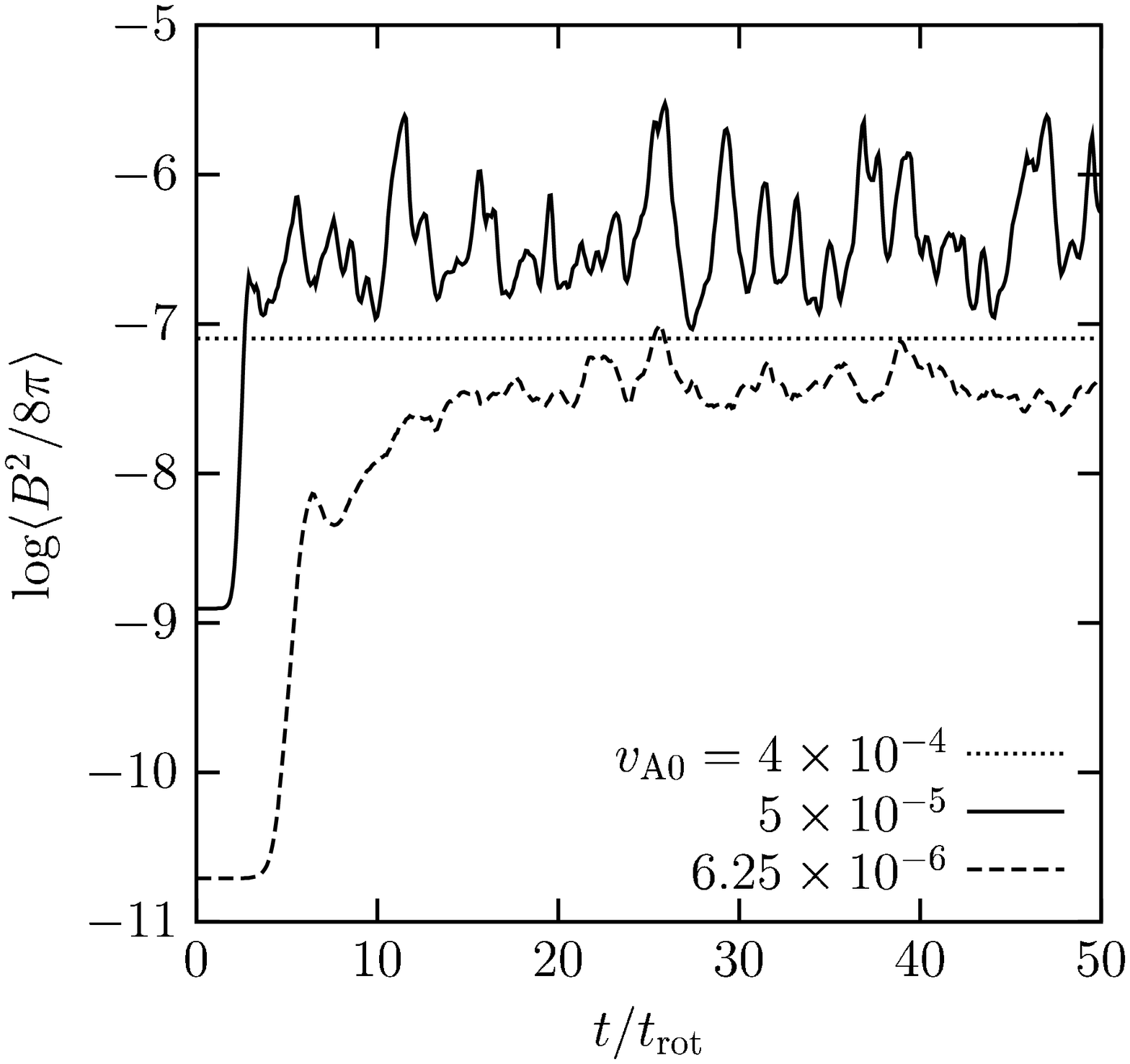}\\
\figcaption
{
Time evolution of the magnetic energy in uniform $B_z$ models Z32
($v_{{\rm A}0} = 6.25 \times 10^{-6}$), Z62 ($v_{{\rm A}0} = 5 \times
10^{-5}$), and Z92 ($v_{{\rm A}0} = 4 \times 10^{-4}$).
The magnetic field in model Z92 is so strong that the MRI wavelength
is longer than the box size.  
The saturation level and time variability in model Z62 are much larger
than in Z32.  
The large spike-shaped variations are due to the recurrent
appearance and breakup of the two-channel flow.
\label{fig:btz}}


\plotone{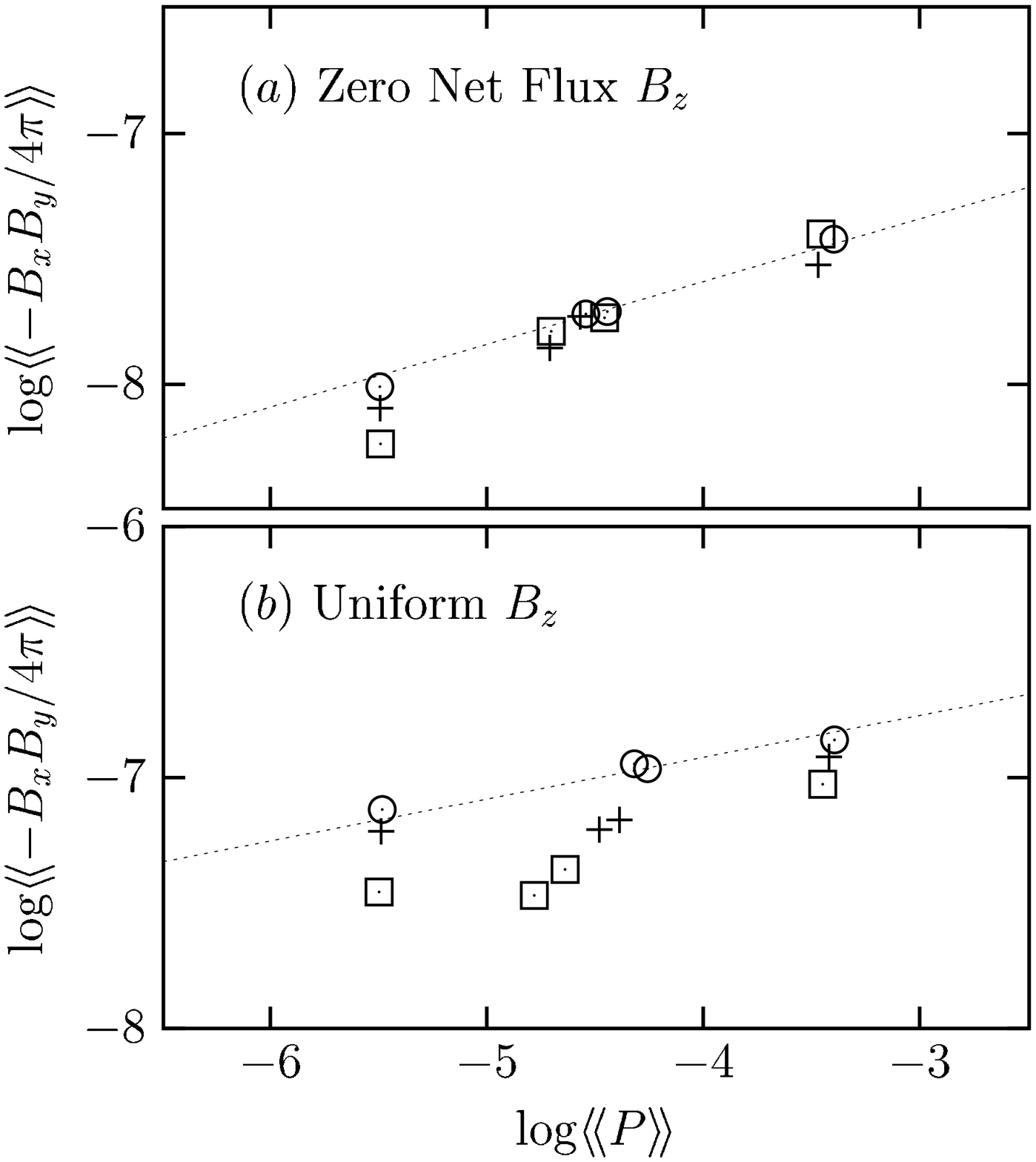}\\
\figcaption
{
Saturation levels of the Maxwell stress in ($a$) zero net flux $B_z$
models (S51, S52, S53, and S52i) and ($b$) uniform $B_z$ models (Z51,
Z52, Z53, and Z52i).  
Circles are from the ideal MHD cases.
Squares are from resistive cases with uniform magnetic diffusivity
$\eta = \eta_0$, and crosses are from cases with an anomalous
diffusivity $\eta = k_0 (v_d - v_{d0})^2$, where $v_d$ is the drift
velocity.
Dotted lines indicate power-laws with exponents ($a$) $q = 1/4$ and
($b$) $q = 1/6$.
\label{fig:eta}}

\clearpage

\plotone{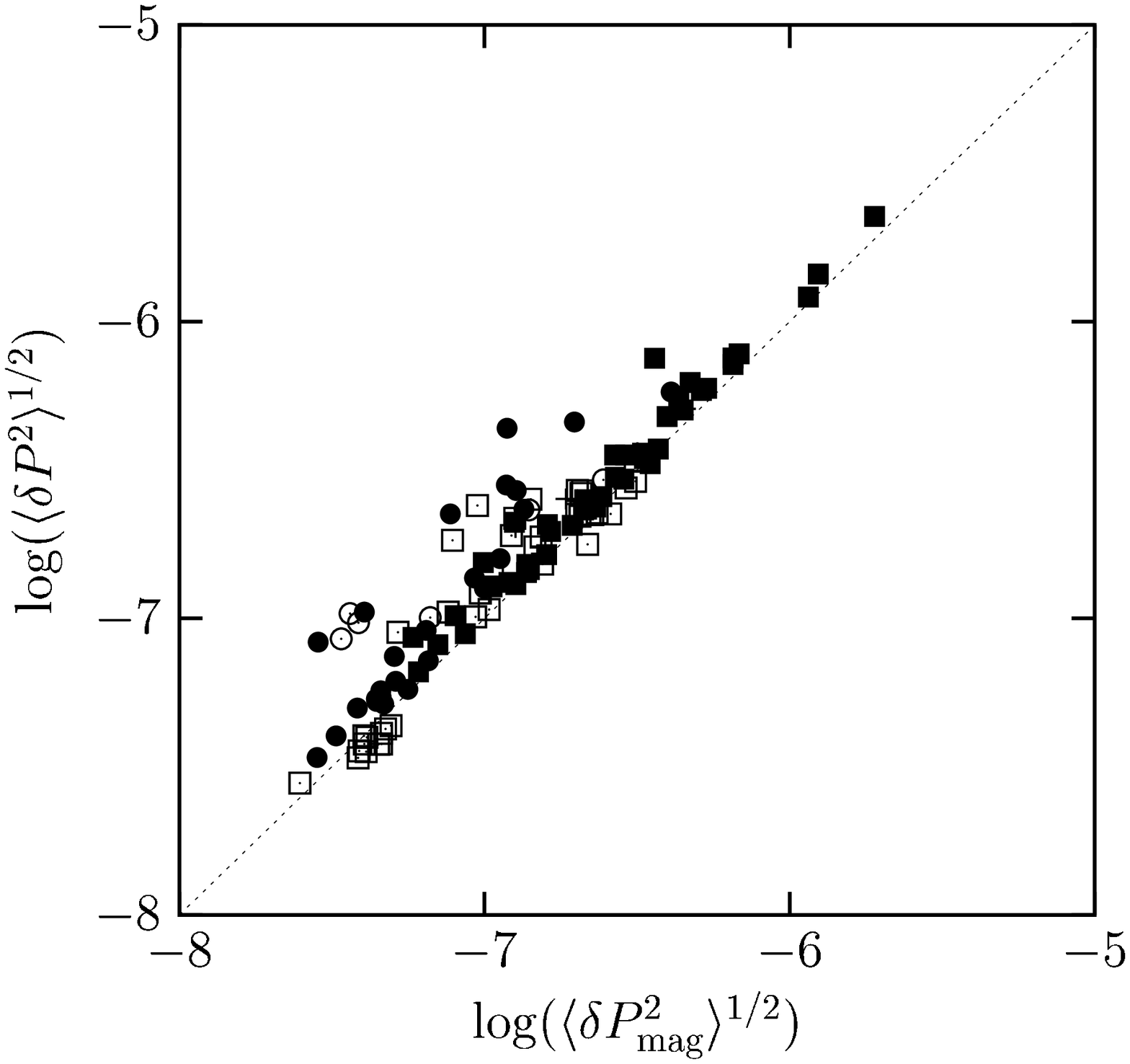}\\
\figcaption
{
The sizes of fluctuations in the gas and magnetic pressures in the
turbulent regime. 
All the models listed in Tables 1 -- 3 are included except for S51i
and S61i (poor resolution), and Z92 (magnetorotationally stable).
For several models (Z51, Z61, Z51i, Z52i, Z61i, and Z62i), the
dispersions of 10 randomly selected snapshots are also plotted.
Open and filled symbols are from ``isothermal'' and adiabatic runs,
respectively.
Circles denote the zero net flux $B_z$ runs and squares are from the
uniform $B_z$ runs.
Crosses indicate 10 snapshots from model Z62i.
\label{fig:dp}}


\plotone{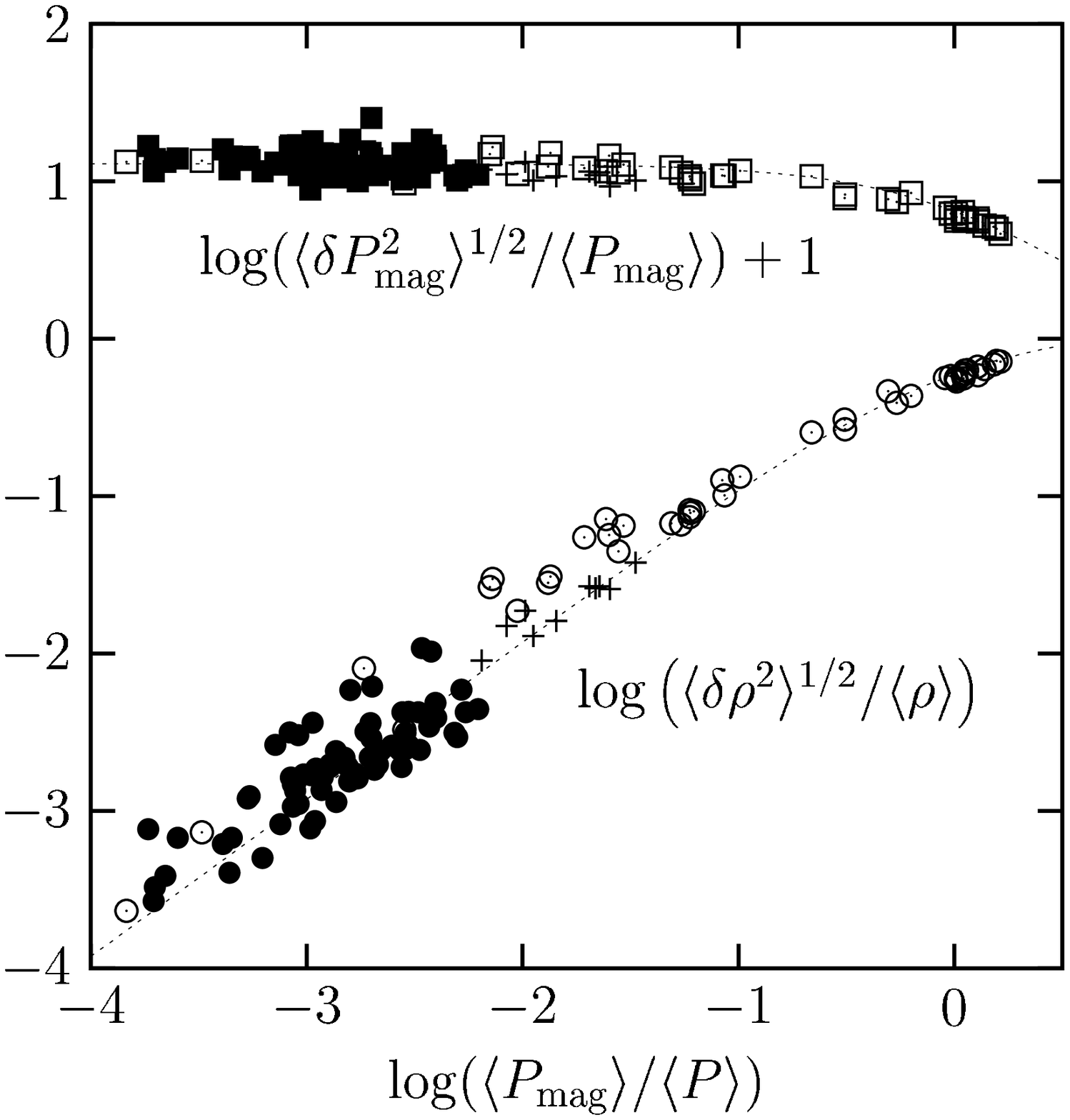}\\
\figcaption
{
The sizes of fluctuations in the density ($circles$) and the magnetic
pressure ($squares$) as functions of the ratio of magnetic to gas
pressure.
Open and filled symbols are from ``isothermal'' and adiabatic runs,
respectively.  
Crosses are from 10 snapshots of model Z62i.
The same models are plotted as in Figure~{\protect{\ref{fig:dp}}}.
All the data are well-fitted by functions 
$\langle \delta P_{\rm mag}^2 \rangle^{1/2} / \langle P_{\rm mag}
\rangle \approx \left( \langle P_{\rm mag} \rangle / \langle P \rangle
+ 1 \right)^{-1}$
and
$\langle \delta \rho^2 \rangle^{1/2}/\langle \rho \rangle \approx 
\left( \langle P \rangle / \langle P_{\rm mag} \rangle + 1
\right)^{-1}$.
\label{fig:rhopm}}

\clearpage

\plotone{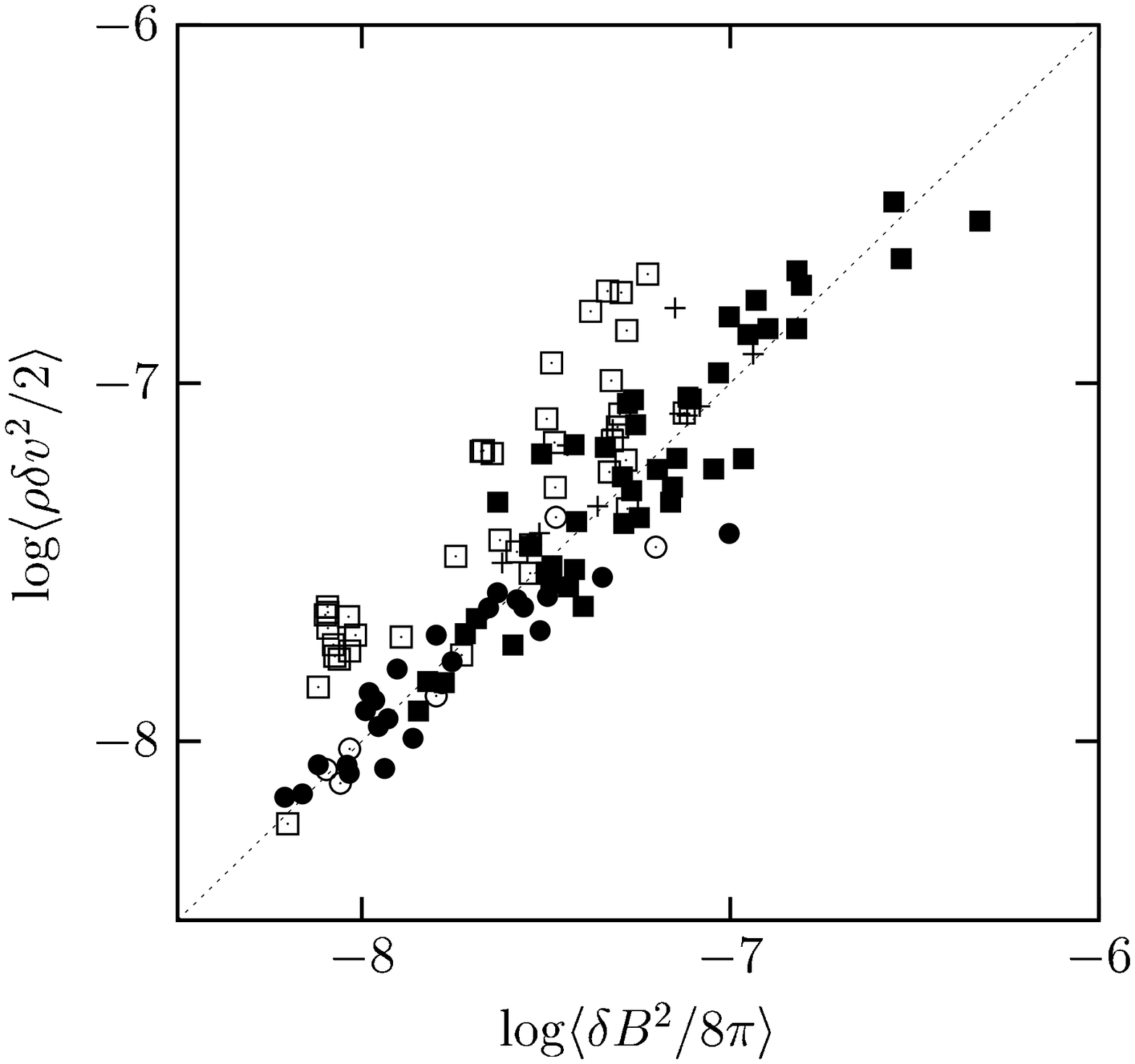}\\
\figcaption
{
The kinetic and magnetic energies of fluctuations.
The models plotted and the meanings of the symbols are identical to
those in Figure~{\protect{\ref{fig:dp}}}.
A dotted line marks equipartition, $x=y$.
\label{fig:dek}}


\plotone{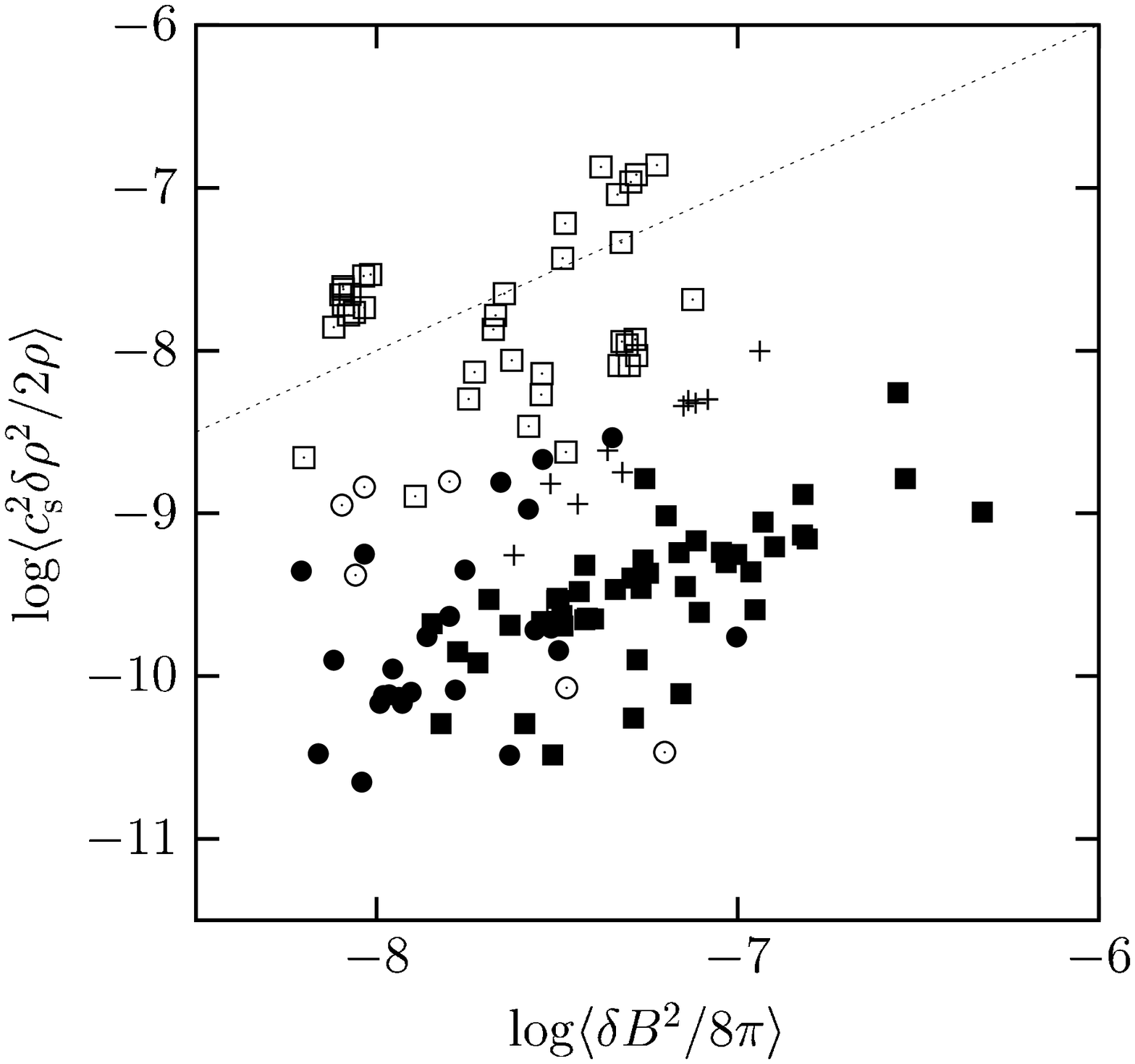}\\
\figcaption
{
The thermal and magnetic energies of fluctuations.
The models plotted and the meanings of the symbols are identical to
those in Figure~{\protect{\ref{fig:dp}}}.
A dotted line marks equipartition, $x=y$.
\label{fig:det}}

\clearpage

\plotone{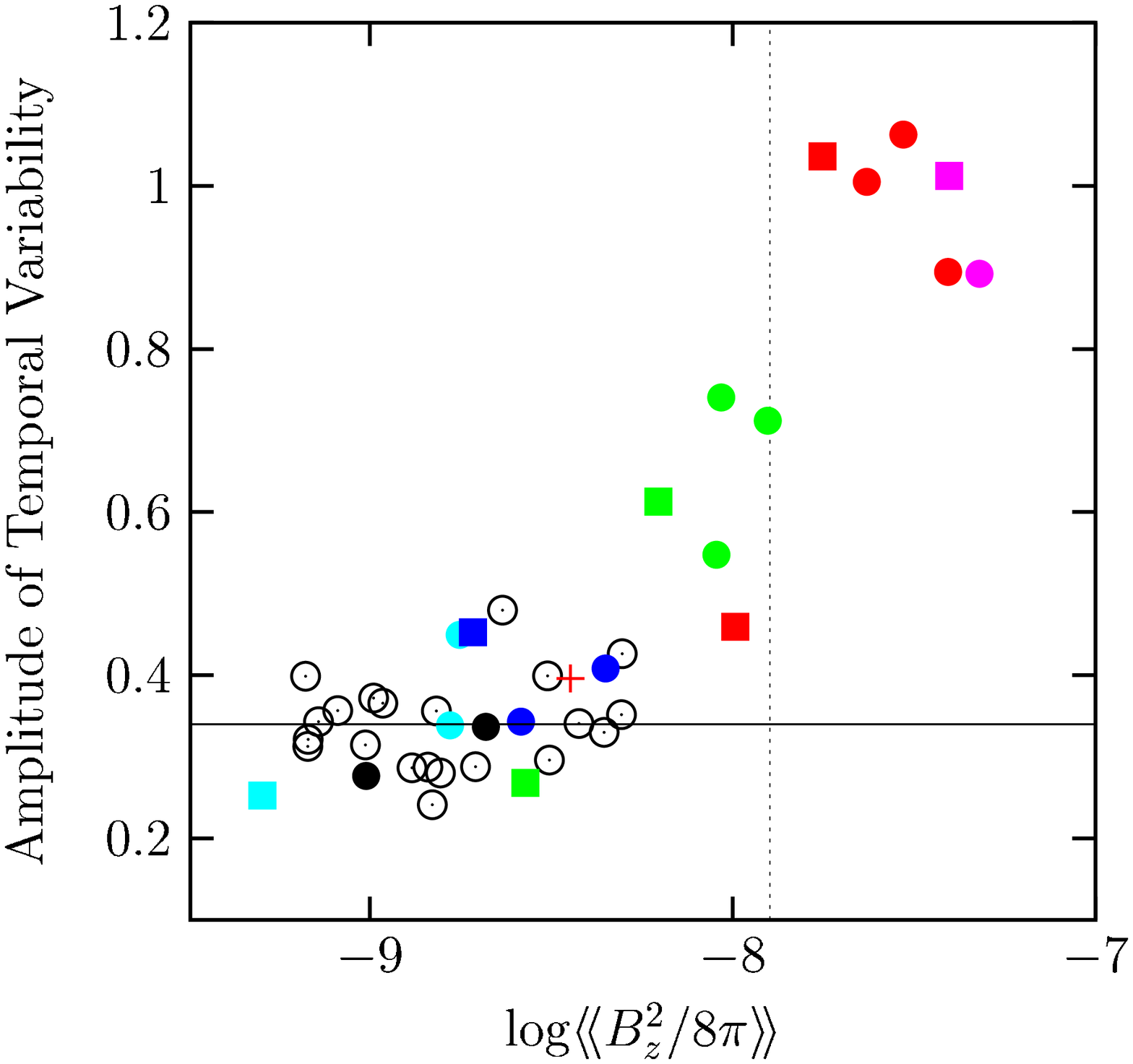}\\
\figcaption
{
Amplitude of temporal variability in the Maxwell stress as a function
of the vertical magnetic energy.
The amplitude is the time-dispersion of the stress normalized by the
time-averaged stress.
Open circles are from the zero net flux $B_z$ models listed in Table 1.
Colored symbols are from uniform $B_z$ models, with meanings as in
Figure~{\protect{\ref{fig:wmpz}}}.  The horizontal solid line shows
the average, 0.34, among the zero net flux runs.
The vertical dotted line indicates where the RMS of the MRI wavelength
$\langle \negthinspace \langle \lambda_{\rm MRI}^2 \rangle
\negthinspace \rangle^{1/2}$ is equal to the vertical size of the
computational domain $L_z$.  When the magnetic field in the nonlinear
regime is strong enough that $\langle \negthinspace \langle
\lambda_{\rm MRI}^2 \rangle \negthinspace \rangle^{1/2} \sim L_z$, the
amplitude of time variations is of order unity.
\label{fig:dispbzall}}

\clearpage

\plotone{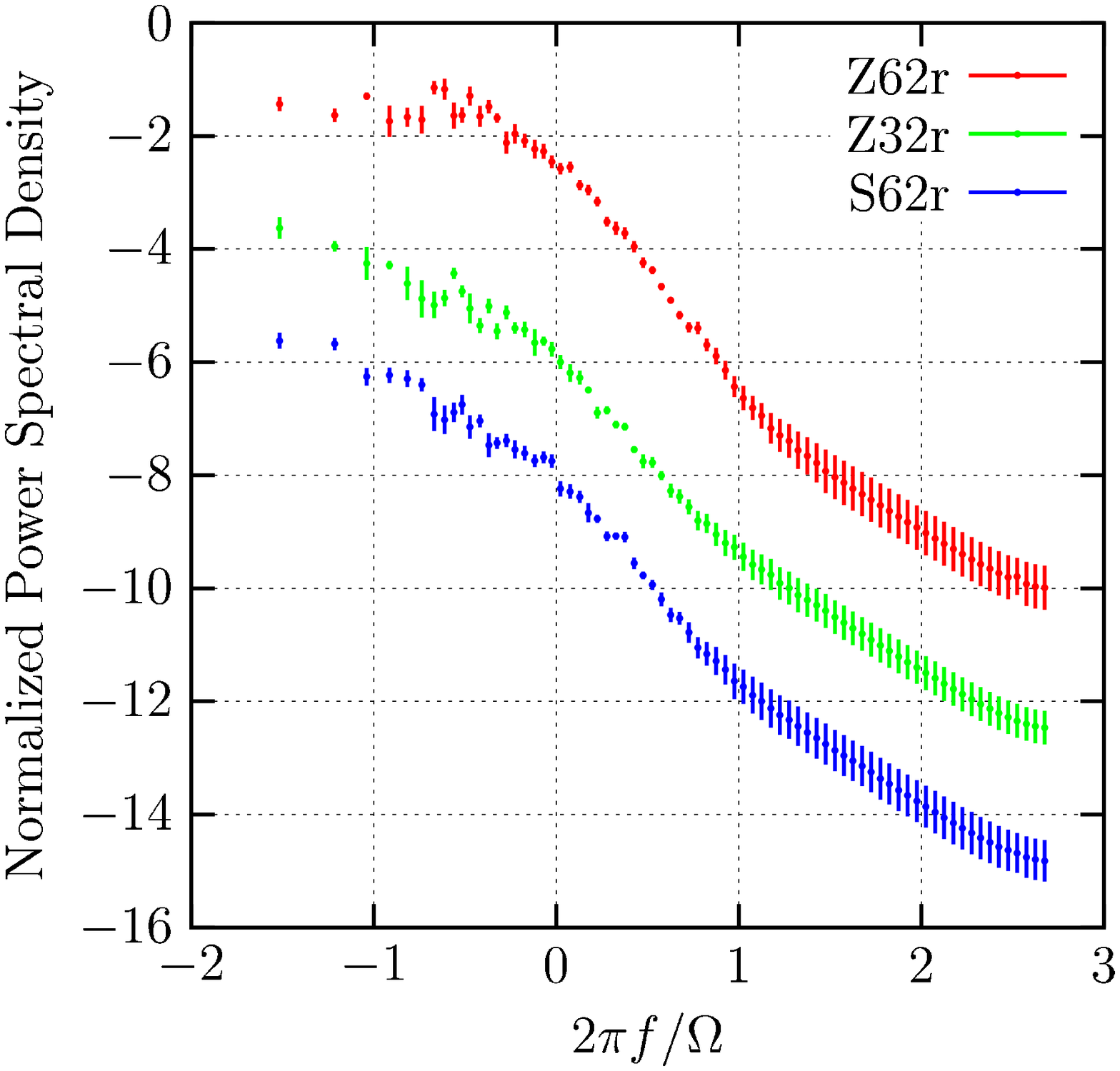}\\
\figcaption
{
Normalized power spectral density for models Z62r (strong uniform
field), Z32r (weak uniform field), and S62r (zero net flux).
The amplitude is normalized by the squared time-average of the stress.
The spectra of Z32r and S62r are divided by
$10^2$ and $10^4$, respectively.
Each spectrum is the average of 8 spectra calculated from 8 segments
in the history data for each model.
The error bars are the standard errors obtained from these 8 spectra.
The power-law indexes of the spectra are listed in Table 5.
\label{fig:npsd}}


\plotone{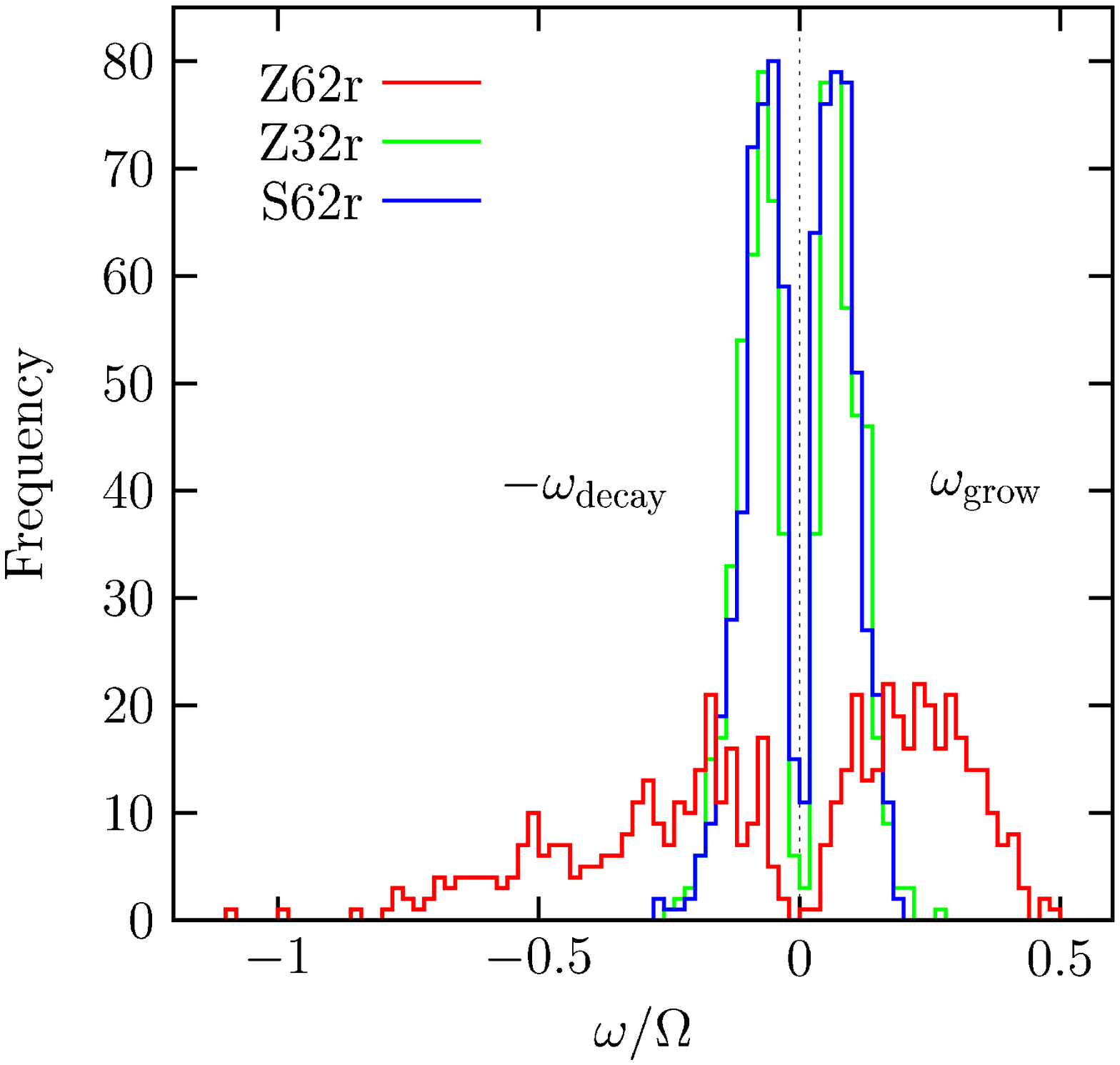}\\
\figcaption
{
Histograms of growth and decay rates in models Z62r (strong uniform
field), Z32r (weak uniform field), and S62r (zero net flux).
The rates are defined by equation (\ref{eqn:omega}).  The distribution
is asymmetric in model Z62r, and symmetric in the other two models.
\label{fig:omegahist}}

\clearpage

\plotone{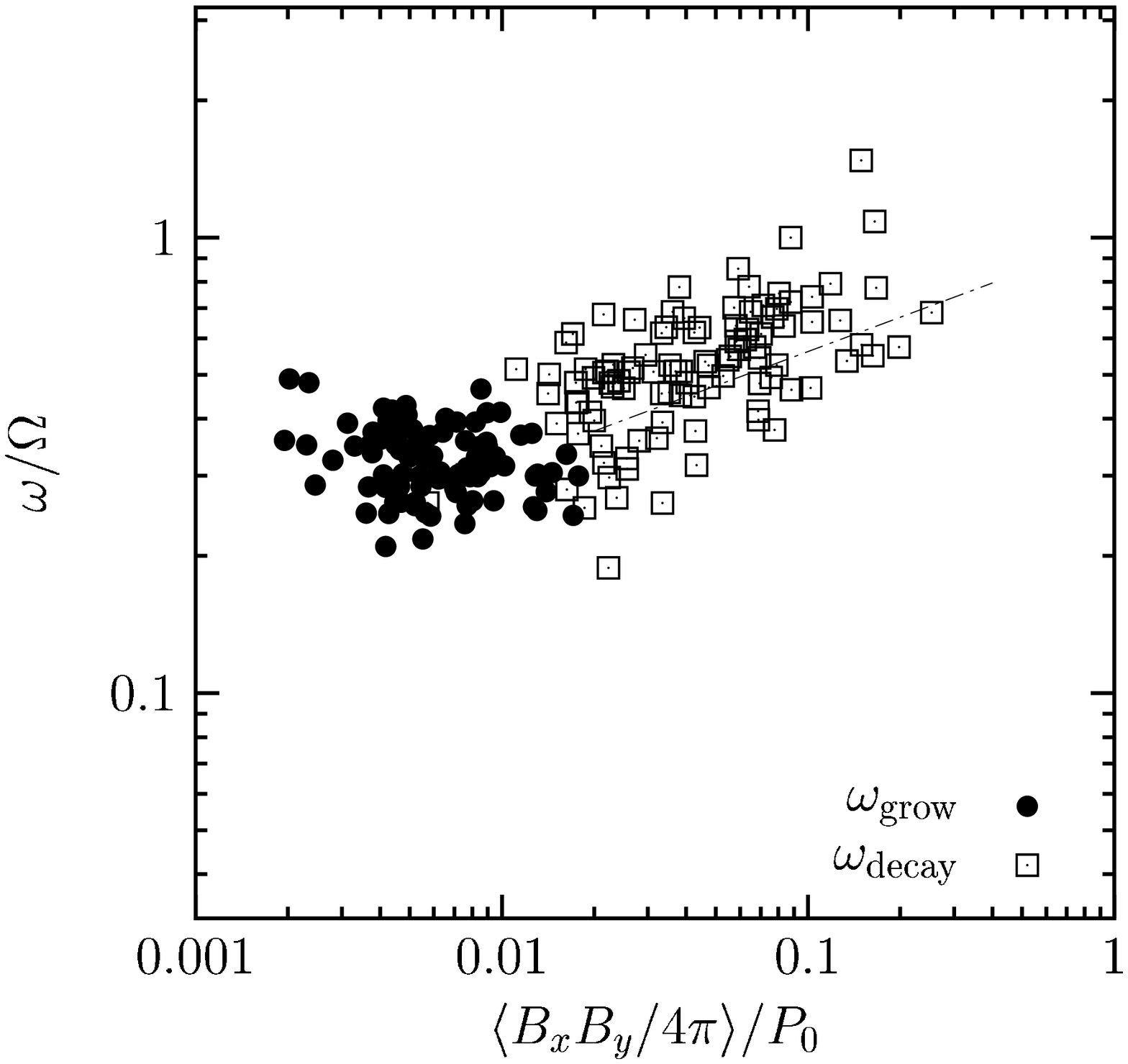}\\
\figcaption
{
The growth ($circles$) and decay ($squares$) rates in model Z62r
as functions of the magnetic stress.
For this figure, data are taken only from time intervals with larger
changes in the stress, $|\ln w_{{\rm ex},j+1} - \ln w_{{\rm
ex},j}| > 1$. 
The growth rate is almost independent of the magnetic stress, but the
decay rate is larger when the magnetic stress and magnetic energy are
larger.  The dot-dashed line indicates the decay rate estimated from a
simple model based on Sweet-Parker type reconnection.
\label{fig:omegawm}}

\end{document}